\documentclass[structabstract]{aa}
\usepackage{graphicx}
\usepackage{txfonts}
\usepackage{natbib}
\bibpunct{(}{)}{;}{a}{}{,}

\usepackage{lscape}
\usepackage{longtable}
\usepackage{amssymb}
\usepackage[latin1]{inputenc}

\def\kms{km~s$^{-1}$}
\def\G23{G023.01$-$00.41}
\def\Jyb{Jy~beam$^{-1}$}

\def\HII{\hbox{H~{\sc ii}}}

\begin{document}
   \title{
   A sub-arcsecond study of the hot molecular core in G023.01$-$00.41\thanks{Based
   on observations carried out with the Submillimeter Array. The
   Submillimeter Array is a joint project between the Smithsonian
   Astrophysical Observatory and the Academia Sinica Institute of Astronomy
   and Astrophysics and is funded by the Smithsonian Institution and the
   Academia Sinica.}
   }


  \author{A. Sanna \inst{1} \and R. Cesaroni \inst{2} \and L. Moscadelli \inst{2} \and Q. Zhang \inst{3} \and K.\,M. Menten \inst{1}
  \and S. Molinari \inst{4} \and A. Caratti~o~Garatti \inst{1} \and J.\,M. De Buizer \inst{5}}
   \offprints{A. Sanna, \email{asanna@mpifr-bonn.mpg.de}}

   \institute{Max-Planck-Institut f\"{u}r Radioastronomie, Auf dem H\"{u}gel 69, 53121 Bonn, Germany
   \and INAF, Osservatorio Astrofisico di Arcetri, Largo E. Fermi 5, 50125 Firenze, Italy
   \and Harvard-Smithsonian Center for Astrophysics, 60 Garden Street, Cambridge, MA 02138, USA
   \and INAF-IFSI, Via Fosso del Cavaliere 100, 00133 Roma, Italy
   \and Stratospheric Observatory for Infrared Astronomy-USRA, NASA Ames Research Center, MS.N232-12, Moffet Field, CA 94035, USA}



  \abstract
   {Searching for disk-outflow systems in massive star-forming regions is a key study to assess the main physical processes in the
   recipe of massive star formation.}
   {We have selected a hot molecular core (HMC) in the high-mass star-forming region \G23, where VLBI multi-epoch observations of water
    and methanol masers have suggested the existence of rotation and expansion within 2000~AU from its center. Our purpose
    is to image the thermal line and continuum emission at millimeter wavelengths to establish the physical parameters and
    velocity field of the gas in the region.}
   {We performed SMA observations at 1.3~mm with both the most extended and compact array configurations, providing sub-arcsecond
    and high sensitivity maps of various molecular lines, including both hot-core and outflow tracers. We also reconstruct the
    spectral energy distribution of the region from millimeter to near infrared wavelengths, using the Herschel/Hi-GAL maps, as well as
    archival data.}
   {From the spectral energy distribution, we derive a bolometric luminosity of $\sim$$\rm4\times10^4~L_\odot$.
    Our interferometric observations reveal that the distribution of dense gas and dust in the HMC is significantly flattened
    and extends up to a radius of 8000~AU from the center of radio continuum and maser emission in the region. The equatorial
    plane of this HMC is strictly perpendicular to the elongation of the collimated bipolar outflow, as imaged on scales of
    $\sim$0.1--0.5~pc in the main CO isotopomers as well as in the SiO(5--4) line. In the innermost HMC regions ($\rm \la1000~AU$),
    the velocity field traced by the CH$_3$CN\,($\rm 12_K-11_K$) line emission shows that molecular gas is both expanding along the
    outflow direction following a Hubble-law, and rotating about the outflow axis, in agreement with the (3-D) velocity field traced
    by methanol masers. The velocity field associated with rotation indicates a dynamical mass of $\sim$$\rm 19~M_{\odot}$ at the
    center of the core. The latter is likely to be concentrated in a single O9.5 ZAMS star, consistent with the estimated bolometric
    luminosity of \G23. The physical properties of the CO\,(2--1) outflow emission, such as its momentum rate
    $6\times10^{-3}$~M$_\odot$~km~s$^{-1}$~yr$^{-1}$ and its outflow rate $2\times10^{-4}$~M$_\odot$~yr$^{-1}$, support our
    estimates of the luminosity (and mass) of the embedded young stellar object.}
  {}


   \maketitle
%

\section{Introduction}

The formation process of high-mass stars ($\rm \ga 10~M_{\odot}$) is still to be established. While
theoretical models have been proposed, only recently substantial progress has
been  made observationally (see, e.g., \citealt{Zinnecker2007}).
It is widely accepted that hot molecular cores (HMCs) are the sites where
massive stars form and it is thus of interest to study these objects at high
angular resolution, through high-density tracers such as rare molecular
species. In the recipe of star formation a crucial role is played by disks
and the associated jets/outflows, which in the case of high-mass stars might
reduce the radiation pressure barrier on the infalling gas (i.e., the so-called flashlight
effect) thus making accretion onto the protostar possible (see, e.g., \citealt{Kuiper2013}).

For these reasons, many observations have been devoted to HMCs and their molecular environment (e.g., \citealt{Cesaroni2007}
and references therein). The success of this type of studies depends mostly on the selection of appropriate targets, as
various facts may hinder the analysis of the velocity field inside the HMCs. For example, too large a distance makes the
linear resolution insufficient to identify any putative disk in the HMC, while a suitable inclination with respect to
the line of sight is needed to detect the velocity gradient due to the outflow expansion and/or disk rotation.
Multi-epoch VLBI observations of maser lines provide us with 3-D information on the velocity field on scales as
small as 10--100~AU and may be used to select promising objects for disk/outflow studies. With this in mind, we have
performed interferometric observations of dense gas and outflow tracers towards the high-mass star-forming region (HMSFR)
\G23. This star forming site harbors a known HMC that we have recently studied by means of high-resolution maser
observations, suggesting the presence of rotation and expansion within 2000~AU from the HMC center \citep{Sanna2010,Sanna2012}.

\G23\ is a HMSFR located at a trigonometric distance of 4.59$^{+0.38}_{-0.33}$~kpc
\citep{Brunthaler2009} which exceeds a bolometric luminosity of $\rm 10^4~L_\odot$ (see \citealt{Sanna2010}
for a recent review).  On a parsec scale, Spitzer/GLIMPSE images of the field reveal bright, ``green fuzzy''
emission at 4.5~$\mu$m, believed to be associated with an intense outflow activity (e.g., \citealt{Cyganowski2009}).
Indeed, a massive $^{12}$CO outflow has been detected with the IRAM interferometer at sub-pc scales (Furuya et al.\,2008).
Towards the center of the molecular outflow, emission from high energy transitions of molecular species such as CH$_3$CN
and NH$_3$ has been detected, indicating the existence of a HMC hosting newly formed massive stars
(\citealt{Codella1997,Furuya2008}). All the strongest maser species known to date have been detected from
the HMC and subsequently imaged with multi-epoch VLBI observations \citep{Sanna2010,Sanna2012}. We hence decided to
perform Submillimeter Array (SMA) observations with sub-arcsec resolution to unveil the gas structure and kinematics
in the HMC and associated outflow on scales ranging from 0.01 to 0.1~pc, with the aim to compare these results to
those obtained from the maser study.

This paper is organized as follows. In Sect.~2, we give the details of our SMA observations. In Sect.~3, we present the results
of the analysis of selected molecular lines for typical HMC and outflow tracers.  We also collect the photometric data at different
IR wavelengthes at the position of \G23, from several Galactic surveys available to date, and reconstruct its spectral
energy distribution. In Sect.~4, we discuss the structure and velocity field of the region from the HMC to the outflow scale
and the nature of the young stellar object(s) (YSOs) embedded in the HMC. Finally, the main results are
summarized in Sect.~5.

\section{Observations and calibration}
\label{smaobs}

The SMA observations toward \G23 were conducted with the very extended (VEX) and compact array configurations
and a single frequency setup in the 230~GHz band. The SMA receiver had two spectral sidebands
each 4~GHz wide; the lower sideband (LSB) was set to include the CH$_3$CN\,(12$_K$--11$_K$) (from 220.747~GHz)
and SiO\,(5--4) (at 217.104~GHz) lines, whereas the upper sideband (USB) was centered at the frequency
of the $^{12}$CO\,(2--1) transition (230.538~GHz). The correlator was configured to an almost uniform spectral resolution of 0.81~MHz
($\sim$1.05~\kms) over the entire bandwidth, except for the spectral windows covering the strongest CH$_3$CN lines ($K$=\,0--6) for which
a resolution of 0.41~MHz (0.52~\kms) was used. The phase center of the observations was the putative position of
the YSO inferred from masers and radio continuum observations (see Table~\ref{tobs}; \citealt{Sanna2010}). Doppler tracking
was performed assuming an LSR velocity of 77.4~\kms, corresponding to the rest velocity of the CH$_3$CN\,(6$_K$--5$_K$)
lines \citep{Furuya2008}. More details on the observations are given in Table~\ref{tobs}.

The visibility data were calibrated using the IDL MIR\footnote{The MIR cookbook is available at the following URL,
https://www.cfa.harvard.edu/~cqi/mircook.html} package and were exported to the MIRIAD format for imaging. The compact
and VEX array configurations had baselines ranging from 7 to 58~k$\lambda$ (corresponding to spatial structures from $29\farcs5$ to $3\farcs5$),
and from 82 to 390~k$\lambda$ (i.e., from $2\farcs5$ to $0\farcs5$). The compact and VEX data were CLEANed
together in order to retrieve information on both spatial scales. Maps were produced by varying the ROBUST parameter to find
a compromise between angular resolution and sensitivity to extended structures. In the following, the resulting synthesized HPBW is reported in
each figure. The continuum data were constructed from line free channels and then subtracted from the \emph{uv}-data (MIRIAD task UVLIN).
All the maps of the outflow tracers were obtained only from the compact configuration data to recover large-scale structures,
with a ROBUST\,0 weighting and a CELLSIZE of $0\farcs3$. Further data analysis was performed with
GILDAS\footnote{The GILDAS software is available at the following URL, http://iram.fr/IRAMFR/GILDAS/}.

\section{Results}

In the following, we describe the results of our SMA observations for selected molecular lines
tracing the hot core and the molecular outflow (see Table~\ref{tlines}),
to analyze the dynamics (from $10^3$~AU to several tenths of pc) of the gas associated
with the radio continuum and maser source observed by \citet{Sanna2010}. We also reconstruct
the spectral energy distribution of \G23\ from the near-IR to the sub-mm, using archival data.

\subsection{Core tracers}\label{core}

In order to analyze the spatial distribution of the HMC emission, besides the dust continuum
emission at 1.3~mm we have selected the strongest molecular lines detected with the VEX
configuration. The overall results of these measurements are summarized in Table~\ref{tcore},
together with the array configuration, weighting used for the CLEANing algorithm, restoring beam
half-power width (HPBW) and position angle, and rms of each map.

\subsubsection{Dust continuum at 1.3~mm}

The first line of Table~\ref{tcore}
gives the integrated continuum flux density of the HMC at 1.3~mm, measured with the compact configuration
of the SMA ($\sim$0.3~Jy). We have not detected any other continuum source within a field of view
of $\sim$1\arcmin\ exceeding a flux density of $\sim$18~m\Jyb\,($3\,\sigma$).
For each array configuration, the continuum emission was fitted with a 2-D
Gaussian using the task IMFIT of MIRIAD. The peak position, intensity, flux densities, and deconvolved angular diameter
obtained from the fit are reported in the last six columns of Table~\ref{tcore}.
A map of the dust continuum emission at 1.3~mm is presented in Figs.~\ref{corepuz}a and~c, superposed on the
ammonia emission map in the (3,3) inversion transition from \citet{Codella1997}, and the map of the
CH$_3$CN\,(12$_2$--11$_2$) line emission obtained by us.

\subsubsection{CH$_3$CN\,(12$_K$--11$_K$) lines}
\label{ch3cn}

The higher spectral resolution used for the CH$_3$CN\,$J$=\,12--11 lines
permits to accurately investigate the velocity field inside the HMC. The analysis presented in
Figs.~\ref{ch3cnspec} and~\ref{velpuz} was performed on the cube obtained from the combined dataset,
cleaned with natural weighting (see Table~\ref{tcore}).

\smallskip
\emph{CH$_3$CN\,(12$_K$--11$_K$) spectral analysis}.
In order to analyze the spectral profile of the methyl cyanide line we use the CH$_3$CN spectrum obtained by
averaging the emission inside the 5\,$\sigma$ contour level of the continuum map (Fig.~\ref{ch3cnspec}). Assuming
that all $K$-components are excited within the same gas, we fitted the $K$=\,0 to 4 components
with 5 Gaussians forced to have same linewidths and with their separations in frequency fixed to the laboratory values.
Note that the $K$=\,0 to 4 components were chosen because they are not significantly affected by line blending with other
molecular species. The free parameters of the fit are the line intensities, the full width at half maximum (FWHM) of the
lines, and the LSR velocity of an arbitrarily chosen $K$ component. The results of the Gaussian fit are reported in
Table~\ref{tch3cn}. The assumptions appear to be justified a posteriori by the excellent agreement between the fit
and the observed spectrum. To derive the physical parameters of the emitting gas, we also fitted the same spectrum
with the XCLASS program (see \citealt{Comito2005} and reference therein), which calculates the line profiles assuming
local thermodynamic equilibrium, for a given molecule column density, source size, and temperature. This time we fitted
all the components up to $K$=\,8. The peak velocity and FWHM were assumed equal to the values obtained from the Gaussian fit.
In Fig.~\ref{ch3cnspec}, we present the CH$_3$CN spectrum and the XCLASS fit (red profile) obtained by varying
the source angular diameter, column density, and temperature. The best-fit parameters are given in
Table~\ref{tch3cn}. The green and blue profiles correspond, respectively, to the CH$_3$CN opacity and the
CH$_3^{13}$CN\,(12$_K$--11$_K$) spectrum computed with XCLASS for the same input parameters used for CH$_3$CN and a
relative abundance $\rm [CH_3CN]/[CH_3^{13}CN]=20$.  The opacity profile demonstrates that the optical depth cannot be
neglected to model the CH$_3$CN emission in \G23, while the abundance ratio is significantly less than the expected
value of 42$\pm$13 (obtained from \citealt{Wilson1994} for a Galactocentric distance of 4.6~kpc), suggesting a
significant enhancement of the $^{13}$C/$^{12}$C ratio in the HMC. One sees some discrepancy
between the fit and the observed spectrum for the $K$=\,0 and~1 lines:  this could be caused by extended emission in the
lower excitation transitions ($K$=\,0 and~1) which is not considered in the model fit, as the latter assumes the same
source size for all $K$ components. Also, we explicitly note that the higher excitation lines of the CH$_3^{13}$CN, which
are not well reproduces by the XCLASS fit, fall within the sensitivity limits of our spectrum.

\smallskip
\emph{CH$_3$CN\,(12$_K$--11$_K$) velocity field}.
In the following, we will assume that the peak velocity of the CH$_3$CN\,(12$_K$--11$_K$) lines of 78.3~\kms\
is the systemic velocity of the HMC ($\rm V_{sys}$).

\begin{figure*}
\centering
\includegraphics [angle=0.0,scale=1.0]{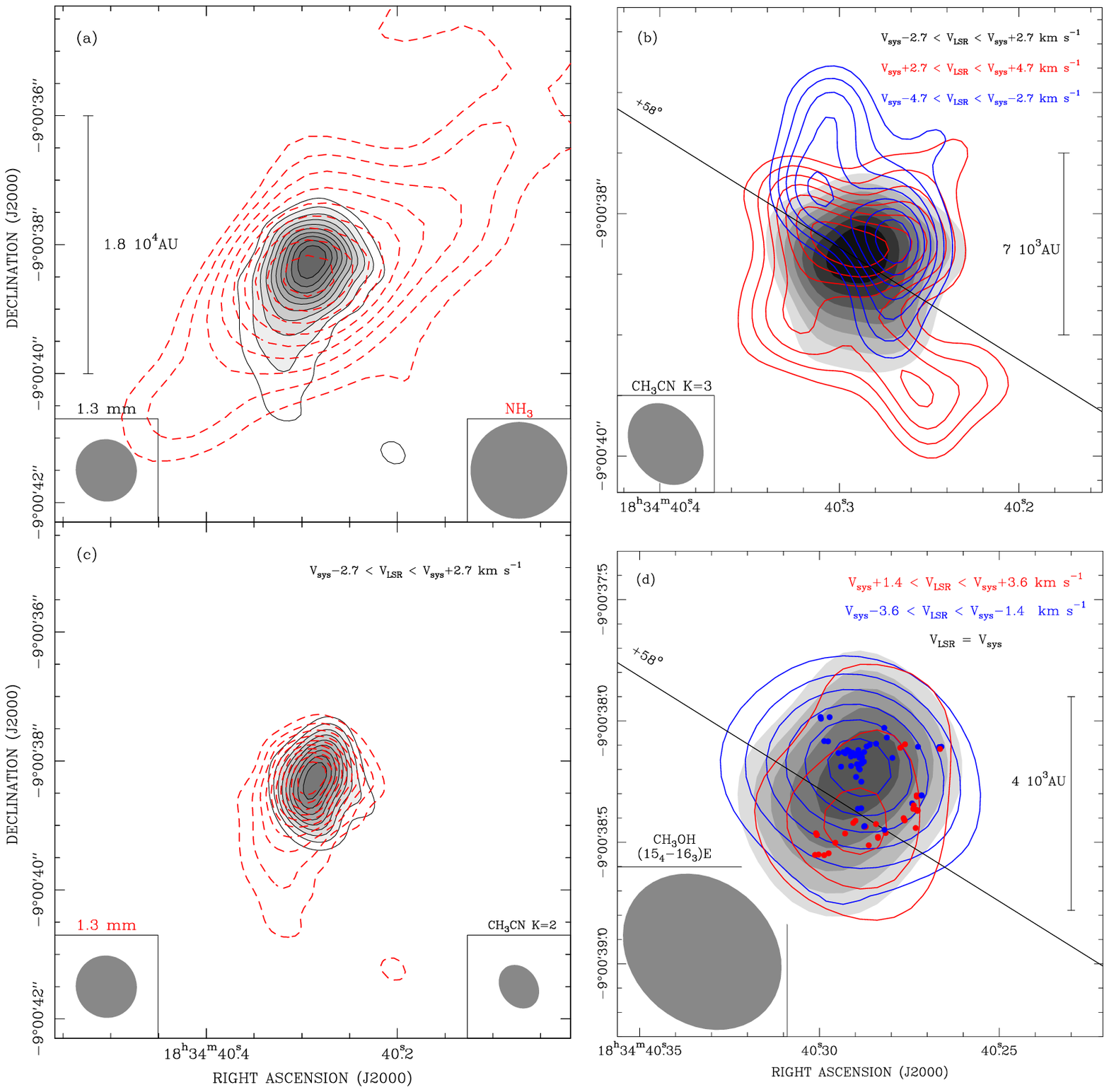}
\caption{Images of the Hot Molecular Core.
\textbf{(a):} SMA map of the 1.3~mm continuum emission from the combination of the Compact and VEX configurations (grayscale), superposed
to the VLA--C NH$_3$\,(3,3) map obtained by \citet[][red dashed contours]{Codella1997}. Contour levels start from 5\,$\sigma$ by 3\,$\sigma$
for the 1.3~mm map (details on Table~\ref{tcore}) and from 3\,$\sigma$ in steps of 1\,$\sigma$ (4~mJy beam$^{-1}$) for the NH$_3$\,(3,3) map.
The SMA and VLA--C synthesized beams are shown in the lower left and right corner, respectively. The linear scale is shown for
comparison with panels~b and~d as well.
\textbf{(b):} Maps of the CH$_3$CN\,(12$_3$--11$_3$) line integrated over its FWHM in three velocity ranges reported on top of the panel
($\rm V_{sys}$=\,78.3~\kms; see Sect.~\ref{core}). Contours start at 3\,$\sigma$ and increase in steps of 1\,$\sigma$. The SMA synthesized
beam is shown in the lower left corner (Table~\ref{tcore}). The NE--SW line shows the outflow direction as described
in Sect.~\ref{outdis}. The f.o.v. is two times smaller than in panel (a).
\textbf{(c):} Similar to panel~(a) with superposed a map of the CH$_3$CN\,(12$_2$--11$_2$) line (grey contours) integrated
over the same bulk emission as panel~(b). The $K=$\,2 line contours start at 3\,$\sigma$ and increase in steps of 1\,$\sigma$. The SMA
synthesized beam is shown in the lower right corner (Table~\ref{tcore}).
\textbf{(d):} Maps of the CH$_3$OH\,($15_4-16_3$)E line emission, the strongest observed in the VEX configuration.
Grey, blue, and red contours are, respectively, maps of emission at the systemic velocity, and in the blue- and red-shifted wings of the
line (the velocity ranges are indicated in the upper right). Grey contours start at 3\,$\sigma$ and increase in steps of 1\,$\sigma$ whereas
the blue and red contours start at a 5\,$\sigma$ level in steps of 2\,$\sigma$. The SMA synthesized beam is shown in the lower left
corner (Table~\ref{tcore}). Blue and red dots mark the positions of the blue- and red-shifted methanol maser spots detected by
\citet{Sanna2010}. The NE--SW line is the same as in panel~(b). The f.o.v. is four times smaller than in panel (a).}
\label{corepuz}
\end{figure*}

\begin{figure}
\centering
\includegraphics [angle=-90.0,scale=0.5]{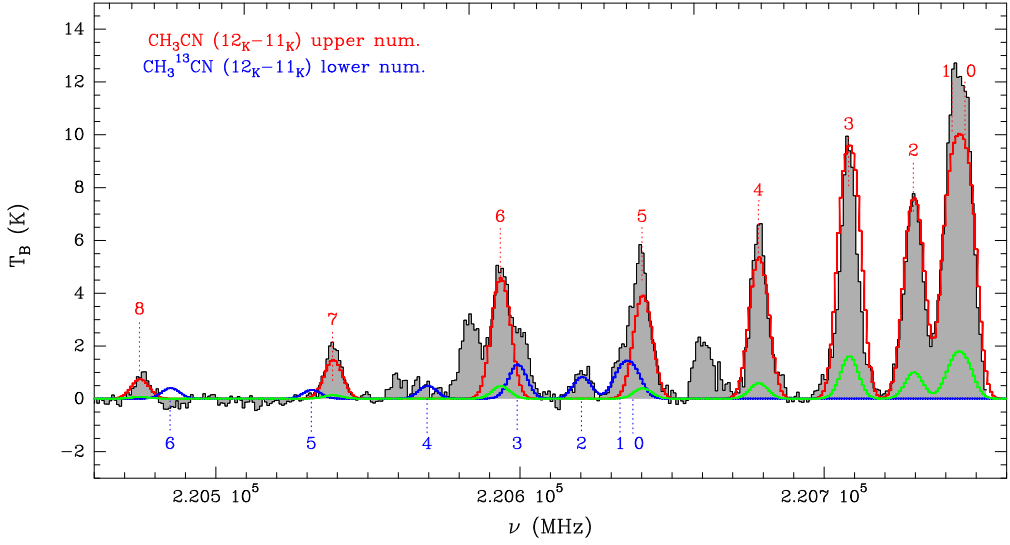}
\caption{Spectra of the CH$_3$CN\,(12$_{K}$--11$_{K}$) lines computed within the $5\,\sigma$ contours of the
dust continuum map (see Sect.~\ref{core}) up to the higher component detectable ($K=$\,8).
The grey profile shows the integrated spectrum and compares the measured profile and the XCLASS synthetic
spectra obtained in LTE approximation for: 1) the CH$_3$CN $K$-ladder (red); 2) the CH$_3^{13}$CN $K$-ladder
(blue); 3) and the opacity profile (green; same scale as the intensity one). The position of the
$K$-ladders for the CH$_3$CN and its isotopomer is marked. Details on Sect.~\ref{core} and Table~\ref{tch3cn}.}
\label{ch3cnspec}
\end{figure}

In Fig.~\ref{corepuz}b, we present a map of the emission from the $K$=\,3 transition within the line FWHM
($\rm V_{sys} \pm 4.7$~\kms), sampled in three velocity intervals. These intervals were selected to analyse the
velocity structure of the core. In particular, a map of the bulk emission was obtained by
averaging the emission over the interval $\rm V_{sys} \pm 2.7$~\kms (see also Fig.~\ref{velpuz}a), while for the maps
of the blue- and red-shifted emission we chose the velocity intervals between $\rm V_{sys} \pm 2.7$~\kms\ and
$\rm V_{sys}\pm{FWHM}/2$. Figure~\ref{corepuz}b shows that the blue and red contours outline a NE--SW bipolar
structure centered on the peak of the bulk emission (grey scale) and consistent with that of the bipolar outflow
mapped by \citet{Furuya2008}. However, on a smaller scale the blue- and red-shifted emissions appear to rotate by
$90\degr$, thus tracing a velocity gradient along the SE--NW direction.

This is better evidenced in Fig.~\ref{velpuz}a, where we present a CH$_3$CN velocity map obtained by
fitting the $K$=\,3 line with a Gaussian in each point of the map. Note that only points with emission above 7\,$\sigma$
have been considered. This map is consistent with that obtained from the line wings, as the velocity gradient seems to
define two directions: in fact, the red-shifted gas is found both to the SE and to the SW, while the red-shifted
emission peaks to the NE and NW. Note that the core appears elongated in the SE--NW direction, as also shown
by the continuum map in Fig.~\ref{corepuz}a. We have computed the position--velocity (\textit{p-v})
plot of the CH$_3$CN\,$K$=\,3 line along the major axis of the core as defined by the dust continuum emission, namely along a
P.A.=\,$-32\degr$ (see Fig.~\ref{velpuz}b). This confirms the existence of a velocity gradient and presents two peaks
symmetrically displaced with respect to the center of the plot, as defined by the position of the continuum peak and the
systemic velocity.

Complementary information is obtained by fitting the emission in each channel (i.e., at fixed velocity)
of the $K$=\,0 to~4 lines, with a 2-D Gaussian. Only channels where the emission was above 7\,$\sigma$
were used. The peak positions orthogonally projected along the two directions with P.A.=\,$58\degr$ and
P.A.=\,$-32\degr$ are plotted as a function of the corresponding velocity in Figs.~\ref{velpuz}c and~d.
The uncertainty of the peak position ($\pm 0.05\arcsec$) has been estimated as the Gaussian fit uncertainty
for a typical SNR of 10 in each channel map.
We then performed a linear fit to the peak distributions; the corresponding slopes ($dv/dx$)
and correlation coefficients ($r$) are, respectively, $-22.4\pm2.4$~\kms\,arcsec$^{-1}$, 0.8 for P.A.=\,$58\degr$
and $-17.2\pm3.3$~\kms\,arcsec$^{-1}$, 0.6 for P.A.=\,$-32\degr$. Note that in Fig.~\ref{velpuz}d
we only fitted the points lying within the FWHM (red and blue dashed lines) of the CH$_3$CN lines.
Out of this range (corresponding to distances greater than about 0\farcs2 from the dust continuum peak)
the points seem to converge towards 0 offset (see discussion in Sect.~\ref{hmcdis}). For comparison, in
Figs.~\ref{velpuz}c and~d we also report the positions of the 6.7~GHz CH$_3$OH masers as measured by \citet{Sanna2010}.
These maser points were not considered in the linear fits.

\subsubsection{CH$_3$OH\,($\rm 15_4-16_3$)E line}\label{meth}

The $15_4-16_3$ methanol line was selected as the highest excitation ($\rm E_{low}\approx 363~K$) and
strongest transition visible in the VEX data, with the aim to compare the thermal methanol emission
with the 6.7~GHz methanol maser emission studied by \citet{Sanna2010}.
The line profile (Fig.~\ref{ch3ohspec}) presents two peaks at about $\rm V_{sys} \pm 2.5$~\kms,
with the blue-shifted peak brighter than the red-shifted one by a factor of almost 2, and a similar
FWHM of about 3~\kms\ for both components. This asymmetry is still present in the low-resolution spectrum
obtained with the compact configuration. The origin of this spectral profile will be discussed in Sect.~\ref{vfieldiscus}.
In Fig.~\ref{corepuz}d, we show the spatial distribution of different velocity components of the methanol gas.
The CH$_3$OH emission was integrated over three velocity ranges:
a map of the blue-shifted emission centered at $\rm V_{sys}-2.5$~\kms\ and integrated over its FWHM,
corresponding to three velocity channels (blue contours); a map of the red-shifted emission centered at
$\rm V_{sys}+2.5$~\kms\ and integrated over the same blue-shifted linewidth (red contours); a channel map of the emission
at the systemic velocity of the core (grey scale). In Table~\ref{tcore} we list the parameters obtained with a 2-D Gaussian fit,
for the methanol emission at the systemic velocity.

\subsection{Outflow tracers}\label{out}

\subsubsection{$^{12}$CO\,(2$-$1) line and isotopomers}\label{coiso}

The CO emission from the $J$=\,2--1 rotational transition is spread over a radius of about $30''$
around the dust continuum peak (Fig.~\ref{outpuz}a, b, and c), consistent with the findings of \citet{Furuya2008}
for the $^{12}$CO(1--0) transition (their Fig.~6a, b, and c). We restrict our analysis to the emission inside the
dashed box in Fig.~\ref{outpuz}a, which better matches the $J$=\,1--0 emission detected with the IRAM interferometer.
Our CO maps reveal the presence of a second source of CO emission (clump ``B'' in Fig.~\ref{outpuz}a) offset by
about 25\arcsec\ to the SW of the HMC. The LSR velocity range of the CO emission from the HMC extends up to
$\pm 30$~\kms\ from $\rm V_{sys}$, as shown in Fig.~\ref{outspec}, whereas at more blue-shifted velocities the
spectrum is affected by the emission from clump~B. In Sect.~\ref{outdis}, we will conclude that clump~B is an
unrelated cloud lying along the line-of-sight and as such it will not be discussed any further.

In order to adopt suitable velocity ranges to identify the blue- and red-shifted outflow emission, we set the inner limits of the
line wings at $\rm V_{sys}\pm{FWHM}/2$, where FWHM is the CH$_3$CN linewidth. This assumption is supported by
the fact that at higher velocities the CH$_3$CN gas tends to align with the outflow direction (see Fig.~\ref{velpuz}d).
These limits correspond to $\pm$4.7~\kms\ from the systemic velocity. As terminal velocities of the flow ($\rm V_{t}$), we have assumed
the velocities where the wing emission falls below the 3\,$\sigma$ level (i.e., at 49~\kms\ and 108~\kms,
respectively). Note that the blue-shifted emission, after decreasing below the 3\,$\sigma$ level at $\rm V_{sys}-18.7$~\kms\
(i.e., at 59.6~\kms), is detected again at higher velocities up to the same (in absolute value) red-shifted terminal velocity
($\rm |V_{t}-V_{sys}|=29.7$~\kms). These terminal velocities correspond to the maximum velocities measured by
\citet{Furuya2008} for the $^{12}$CO\,(1--0) and $^{13}$CO\,(1--0) line emission. In Fig.~\ref{outpuz}c, we show the
bulk C$^{18}$O\,(2--1) line emission averaged over the FWHM of the CH$_3$CN lines, for comparison with
Fig.~6c of \citet{Furuya2008}. In Fig.~\ref{12cube} and~\ref{13cube}, channel maps of the blue- and red-shifted wings of
the $^{12}$CO\,(2--1) and $^{13}$CO\,(2--1) emission are shown in pairs of channels equally offset (in absolute value)
from the systemic velocity.

The physical properties of the $^{12}$CO\,(2--1) outflow are reported in Table~\ref{tout} for the blue- and red-shifted
lobes separately (following, e.g., \citealt{Cabrit1990,Beuther2002}). The length of each CO lobe was calculated from
the peak of the dust continuum map to the farthest CO emission detected along the NE--SW direction (plus signs in
Fig.~\ref{outpuz}a).  Since the spatial distribution of the CO gas shows  a clumpy morphology (e.g.,
Fig.~\ref{12cube}), and the maximum outflow velocity does not increase linearly as a function of the distance from the
HMC, the outflow properties are estimated for a median velocity of 15~\kms. For calculating the CO outflow mass, we
made use of the first equation in Sect.~3.2.2 of \citet{Qiu2009}, assuming an excitation temperature of 30~K
and a $^{12}$CO abundance relative to H$_2$ of $10^{-4}$. The excitation temperature was adopted for comparison with
the SiO outflow properties (see below). Note that decreasing the excitation temperature by a factor 2, for consistency
with $\rm T_{ex}$=\,15~K adopted by \citet{Furuya2008}, reduces the outflow mass estimates by less than a factor 1.3.
We applied a constant correction for the $^{12}$CO opacity ($\tau_{12}$) of 5.3, derived from the second equation in
Sect.~3.2.2 of \citet{Qiu2009} for a $^{12}$CO/$^{13}$CO abundance ratio of 42 (from \citealt{Wilson1994}).
This average optical depth was derived from the flux density ratio between the red-shifted outflow lobes of
Fig.~\ref{outpuz}a and~b (within the 3\,$\sigma$ level of the $^{12}$CO emission), where the $^{12}$CO and $^{13}$CO
gas components are well matched in position. Note that, assuming the same $^{12}$C/$^{13}$C abundance ratio ($\sim$20)
as that obtained for the HMC from the CH$_3$CN and CH$_3^{13}$CN molecules, the outflow mass estimates
are reduced by a factor 2.

In Fig.~\ref{outpv}a, we also present a \textit{p-v} plot of the $^{12}$CO\,(2--1) emission along a position angle
of $+58\degr$ passing through the peak of the continuum emission. The CO emission was averaged over a strip about
$2\arcsec$ wide from each side of the cut, in order to include most of the emission from the outflow lobes.
We will further comment on the outflow from \G23\ in Sect.~\ref{outdis}.

\subsubsection{SiO\,(5$-$4) line}
\label{sio}

The emission from the $J$=\,5--4 line of the SiO molecule arises from a region of 6\farcs7 (0.15~pc)
around the dust continuum peak and covers the velocity range $V_{\rm sys} \pm 23.1$~\kms\
(Fig.~\ref{outpuz}d and \ref{outspec}). These limits correspond to the velocities of the channel maps where the emission from both the
blue- and red-shifted wings drops below the $3\,\sigma$ level. The SiO molecule is almost exclusively associated
with high velocity gas and its gas-phase abundance is greatly enhanced by strong shocks passing through
dense molecular gas (and disrupting dust grains). Therefore, we integrated the SiO emission starting from the
spectral channels next to the systemic velocity (Fig.~\ref{outpuz}d).

The physical properties of the SiO outflow are reported in Table~\ref{tout} for the blue- and red-shifted lobes separately.
The length of the SiO lobes was calculated from the peak of the dust continuum map to the farthest (3\,$\sigma$ contours)
SiO emission along a direction with P.A.=$+58\degr$  (black line in Fig.~\ref{outpuz}d). The outflow velocity used for these
estimates is the maximum line-of-sight velocity observed (i.e., 23.1~\kms). For the calculation of the SiO outflow mass,
we used Eq.~(1) of \citet{Gibb2004}, assuming an excitation temperature equal to the energy of the $J$=\,5 level (i.e., 31~K),
and an SiO abundance relative to H$_2$ of $2 \times 10^{-9}$ \citep{Gibb2007}. It is worth noting that both the SiO outflow
velocity and the excitation temperature are likely lower limits as well as the inferred quantities. Also, we explicitly
note that the SiO abundance is still affected by large uncertainties (e.g., \citealt{Gibb2007}). In Fig.~\ref{outpv}b,
we present a \textit{p-v} plot of the SiO\,(5--4) emission close to the HMC  along a cut at P.A.=$+58\degr$, for comparison
with the velocity gradient observed in CH$_3$CN. In Fig.~\ref{outpv}a, the
same \textit{p-v} plot (white contours) is superposed to the one inferred
from the $^{12}$CO\,(2--1) line.

We conclude that the SiO outflow morphology and physical parameters are consistent with those obtained from the CO isotopomers,
within the uncertainties.

\subsection{Spectral Energy Distribution}\label{SED}

Table~\ref{tsed} lists the flux densities from 3.4~$\mu$m to 1.1~mm toward \G23\ obtained from the Herschel
Infrared Galactic Plane Survey (Hi-GAL; \citealt{Molinari2010}) images and the following archival data:
Wide-field Infrared Survey Explorer (WISE); Galactic Legacy Infrared Mid-Plane Survey
Extraordinaire (GLIMPSE; \citealt{Benjamin2003}); MSX \citep{Egan2003}; MIPSGAL \citep{Carey2009};
APEX Telescope Large Area Survey of the Galaxy (ATLASGAL; \citealt{Schuller2009,Csengeri2014}); Bolocam Galactic Plane
Survey (BGPS; \citealt{Drosback2008}). For the Herschel/Hi-GAL data, the most recent release was used (obtained in
2011 July with HIPE version 7.0.0), resulting from data reduction with ROMAGAL software \citep{Traficante2011},
where the most relevant image artifacts have been removed by a weighted post-processing of the generalized least square
maps \citep{Piazzo2012}.
In Fig.~\ref{irplots}a and~b, we show a Herschel picture of the HMSFR \G23\ obtained with the PACS camera at 160~$\mu$m and 70~$\mu$m,
which indicates that the emission at infrared wavelengths peaks towards the HMC. In Fig.~\ref{irplots}c, we also show that the
$^{12}$CO\,(2--1) outflow emission discussed in Sect.~\ref{coiso} is centered on the peak position of the infrared emission and the
dusty clump imaged with the APEX telescope at 870~$\mu$m. We hence assume the HMC itself to be responsible for most of the far-IR emission
from the region and to dominate the bolometric luminosity.

The flux densities were estimated by integrating the emission inside a suitable region chosen to minimize the contamination
from nearby cluster members  (visible in Fig.~1 of \citealt{Testi1998} and Fig.~1 of \citealt{Furuya2008}). The radius of
this region was $\sim 0.6\arcmin$ (dashed circle in  Fig.~\ref{irplots}b) in all cases except the GLIMPSE images, for which
a radius of $\sim$6\arcsec\ was used, as the angular resolution is significantly higher than in the other images.
For the flux calculation, we also subtracted a mean value of the background emission. In each image, the latter was estimated
by performing a number of 1-D cuts across the peak and measuring the value of the flux at the intersections between such cuts
and the border of the selected region over which the integration was performed. Clearly, the background level estimated in this
way depends on the direction of the cut, but we find that the corresponding uncertainty on the flux density estimate
does not exceed 20\%.

We have fitted the SED with the radiative transfer model developed by \citet[][by using the SED fitting tool available at http://caravan.astro.wisc.edu/protostars/]{Robitaille2007},
which assumes a pre-main-sequence star with a circumstellar disk, embedded in an infalling flattened envelope with outflow
cavities, and allows derivation of a number of physical parameters. While this model cannot take into account the details
of the ongoing star formation toward \G23\ (e.g., \citealt{Offner2012}), the fit should be sufficiently good to give
reliable estimates of \emph{integrated} quantities such as the luminosity and mass of the clump (solid black
profile in Fig.~\ref{sed}). Fixing the distance at 4.6~kpc, the best-fit values are $\rm 3.9\times10^4~L_\odot$ and
$\rm 1.4\times10^3~M_{\odot}$, respectively, with an interstellar extinction of $\rm A_V$=\,87~mag.
According to \citet{Offner2012}, we caution that the mass value may be significantly overestimated,
because the model does not take properly into account the possible presence of a cold component of the clump.

\section{Discussion}\label{discussion}

In the following section, we will discuss the characteristics of the hot molecular core and associated outflow
in \G23, in the light of two features revealed by our SMA observations on scales from $1\arcsec$ to $0\farcs1$: 1)
the flattened morphology of the core, perpendicular to the outflow direction; and 2) the velocity field revealed
through the CH$_3$CN emission, which appears to outline two mutually orthogonal directions at sub-arcsecond scales
($\lesssim 3\times10^3$~AU). We also remind that the HMC is known to contain a compact source of radio continuum
emission and a number of CH$_3$OH and H$_2$O maser features (\citealt{Sanna2010}, their Fig.~4).

\subsection{HMC structure}\label{hmcdis}

It is interesting to note that the deconvolved size of the core derived from different tracers and with different
angular resolutions has a fairly constant ratio of $\sim$2 between the major and minor axes (Table~\ref{tcore}).
This result lends support to the existence of a real structure whose characteristics do not depend on the tracer
observed. Assuming a circular disk symmetry, this implies that the rotation axis is inclined by $30\degr$
with respect to the plane of the sky.

At the largest scale imaged with the compact configuration, the dust emission traces a region as large as
$3\farcs4$ (see Table~\ref{tcore}), corresponding to a radius of about 8000~AU from the center of the core. The
emission from the (3,3) inversion transition of ammonia arises from a similar region, with a deconvolved angular
diameter of $3\farcs3$ (Fig.~\ref{corepuz}a; see also \citealt{Codella1997}). The 1.3~mm flux is about ten times the
flux recovered at 3~mm by \citet{Furuya2008} within a similar angular extent ($\rm \la 3\arcsec$), that implies a
spectral slope in the mm domain of $\gamma = 3.1 \pm 0.4$ and a dust emissivity exponent $\beta \sim 1$.
At the improved resolution of the combined configuration (Compact+VEX), both the dust continuum emission and the
bulk emission of the CH$_3$CN\,$K$=\,2 line present a more compact structure with a major axis of $\sim1\farcs3$,
corresponding to a radius of $\sim$3000~AU (Fig.~\ref{corepuz}c). We also note that
the size of the CH$_3$CN emission inferred from the XCLASS program ($\sim1\farcs2$; see Table~\ref{tch3cn}) is in good
agreement with that measured from the maps, lending support to the reliability of the fit.
Information about the most compact structure that can be imaged with the SMA is obtained
from the CH$_3$OH\,(15$_4$--16$_3$)\,E transition, whose excitation energy is more than four times higher than
that of the CH$_3$CN\,(12$_2$--11$_2$) line. This methanol line unveils a flattened core, with a radius of
$\sim$2000~AU (Fig.~\ref{corepuz}d).

An interesting feature of the emission from the HMC core, is that the size of it appears to depend on the
tracer: higher excitation lines appear to arise from smaller regions. This trend may be due to increasing
gas temperature towards the core center, which in turn would imply the presence of an embedded heating
source. However, when comparing different tracers one must take into account that a variation of
optical depth may also produce the same effect, with the more opaque lines tracing larger regions.
For instance, in Table~\ref{tcore} we show that the deconvolved size
of the CH$_3$CN\ emission in the $K$=\,7 component ($E_{\rm low}\approx 408~K$) is a factor $\sim 2$ less than
that of the $K$=\,2 and~3 components ($E_{\rm low}\approx 87~K$ and 123~K, respectively). But for these lines
the opacity increases from the $K$=7 to the $K$=2 components (Fig.~\ref{ch3cnspec}), which might also explain
the corresponding increase in size. However, it is possible to demonstrate that for an homogenous sphere the
ratio between the FWHM of the source in the optically thick and thin limits is $2/\sqrt{3}$, significantly less
than the factor 2 measured by us. We thus conclude that in all likelihood the observed decrease in size is, at
least in part, a temperature effect, witnessing the presence of one or more deeply embedded YSOs.

\subsubsection{Velocity field of the HMC}\label{vfieldiscus}

As noted in Sect.~\ref{meth}, the methanol line from the HMC shows a double-peak profile, with blue-shifted emission
stronger than the red-shifted one. In Fig.~\ref{corepuz}d, we show that such a feature is the result of two partly
overlapping components that peak at both sides of the outflow axis. The position of the outflow axis was centered on
the dust continuum peak (see Sect.~\ref{out}). Comparison with the spectrum of the CH$_3$OH maser at
6.7~GHz toward the HMC shows that also the maser lines consist of two groups \citep[their Fig.~1]{Sanna2010},
lying at the same red- and blue-shifted velocities as the CH$_3$OH thermal emission. In the VLBI maps these
two groups of maser lines correspond to two distinct clusters of spots coincident with the blue- and red-shifted peaks of
the thermal emission (blue and red dots in Fig.~\ref{corepuz}d). \citet{Sanna2010} show that the proper motions
of individual masing cloudlets describe two different velocity fields: rotation
about an axis with P.A.$\simeq$60\degr, and expansion along the direction of this axis.
Assuming centrifugal equilibrium, the dynamical mass estimated from the rotational component of the
methanol masers is $\rm \sim20~M_\odot$.
The similarity between the spectrum and spatial distribution of the thermal and maser emission suggests that both
methanol transitions could be excited in the same physical environment.

\begin{figure*}
\centering
\includegraphics [angle=0.0,scale=1.1]{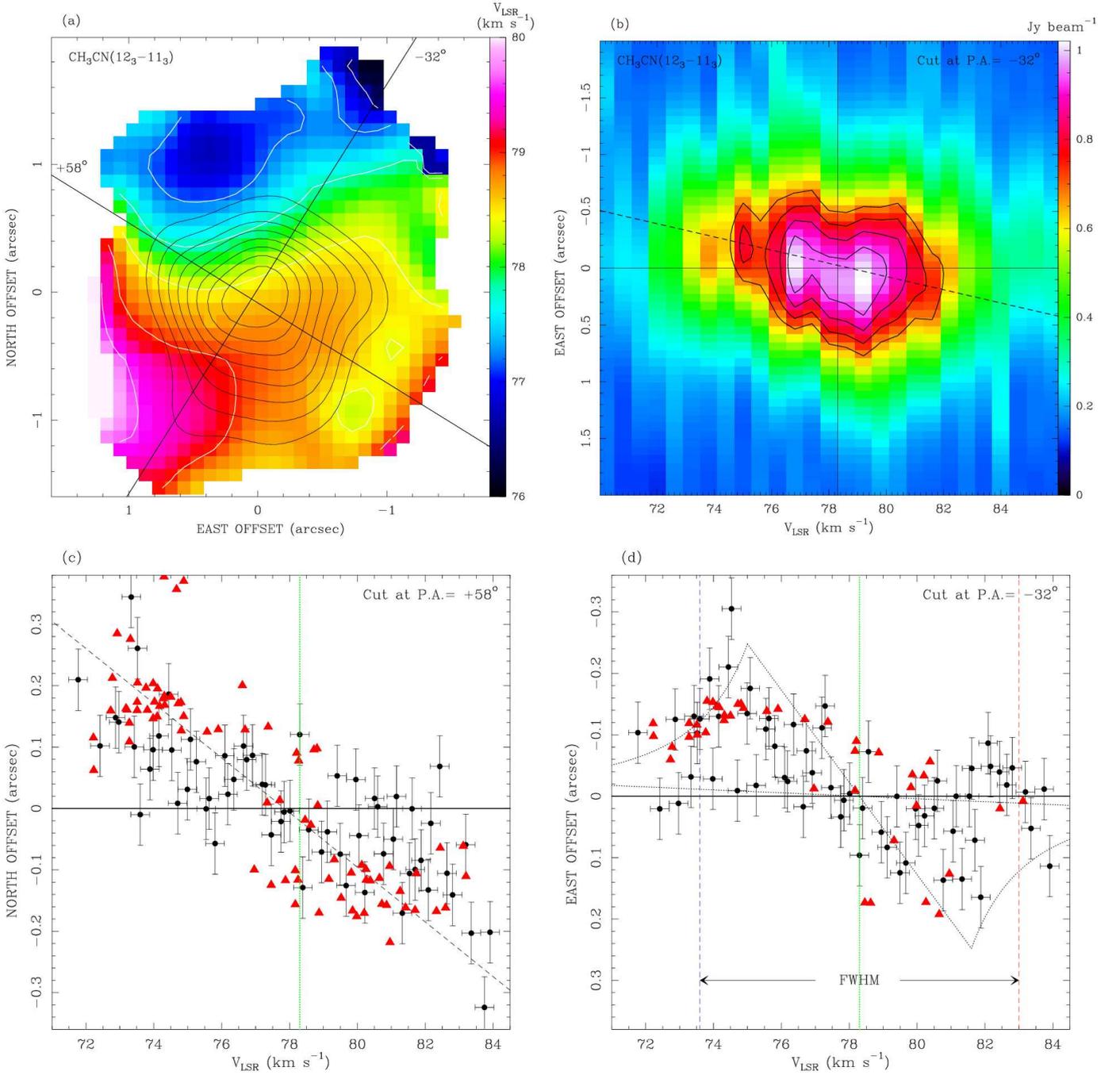}
\caption{Velocity pattern of the inner core from the CH$_3$CN\,(12--11) lines. In all panels the vertical and
horizontal lines mark, respectively, the systemic velocity of the core and the position of the dust continuum peak.
\textbf{(a):} Map of the CH$_3$CN\,($12_3-11_3$) line velocity (color scale).
Offsets are measured with respect to the phase center. Superimposed are the contours of the bulk emission from
Fig.~\ref{corepuz}b (black) and the directions (straight lines), passing through the continuum peak, along which
the \textit{p-v} cuts presented in the following panels have been computed.
\textbf{(b):} \textit{p-v} cut of the CH$_3$CN\,($12_3-11_3$) line along the major axis of the elongated HMC ($-32\degr$);
east offsets are measured along this cut.
Contours start at 60\% of the peak emission and increase by 10\% steps; colors are drawn according to the wedge on the
right side. The dashed line represents the linear fit performed in panel~(d) within the FWHM of the CH$_3$CN\,($12_K-11_K$) lines.
\textbf{(c):} \textit{p-v} distribution of the peaks of the CH$_3$CN emission (black dots) at different velocities (only the
$K$=0 to 4 components have been used) along the outflow direction ($+58\degr$); north offsets are measured along this cut.
Error bars indicate the spectral resolution
of the CH$_3$CN\,(12--11) observations and the uncertainty of the Gaussian fits in the maps. The black dashed line is the best
linear fit to the peak distribution, which has a slope $dv/dx=-22.4\pm2.4$~\kms\,arcsec$^{-1}$, and a correlation coefficient
of $r$=0.8. The red triangles represent the CH$_3$OH maser spots detected by \citet{Sanna2010}.
\textbf{(d):} Same as panel~(c) for the cut along the major axis of the elongated HMC (P.A.=$-32\degr$); east offsets are
measured along this cut. The vertical dashed lines mark the limits of the FWHM of the CH$_3$CN\,(12--11) lines. The best linear
fit (not shown in this figure) to the points within the line FWHM is obtained for $dv/dx=-17.2\pm3.3$~\kms\,arcsec$^{-1}$,
with $r$=0.6. The dotted pattern encompasses the region where emission is expected from a Keplerian disk rotating about
a $\rm 19~M_{\odot}$ star. Here, the red triangles mark only the CH$_3$OH maser spots with a distance of less than
0.25~arcsec from the dust continuum peak.}
\label{velpuz}
\end{figure*}

\begin{figure}
\centering
\includegraphics [angle=0.0,scale=0.4]{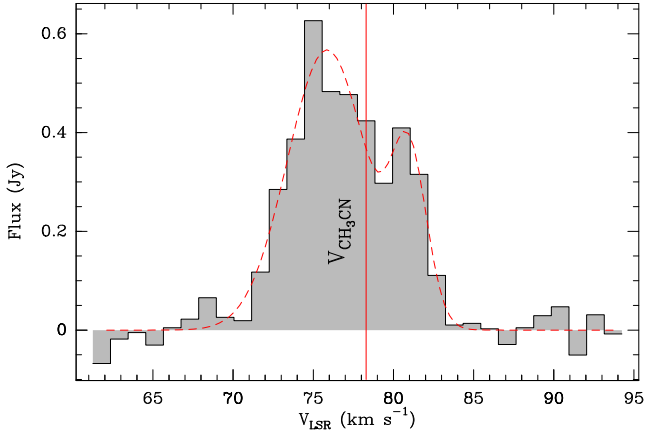}
\caption{ Spectrum of the CH$_3$OH\,($15_4-16_3$)E line emission presented in Fig.~\ref{corepuz}d. The red dashed line on
the spectrum draws a double Gaussian fitting to the blue- and red-shifted wings of the CH$_3$OH profile, symmetric with
respect to the systemic velocity of the CH$_3$CN lines (red vertical line).}
\label{ch3ohspec}
\end{figure}

\begin{figure*}
\centering
\includegraphics [angle=0.0,scale=1.0]{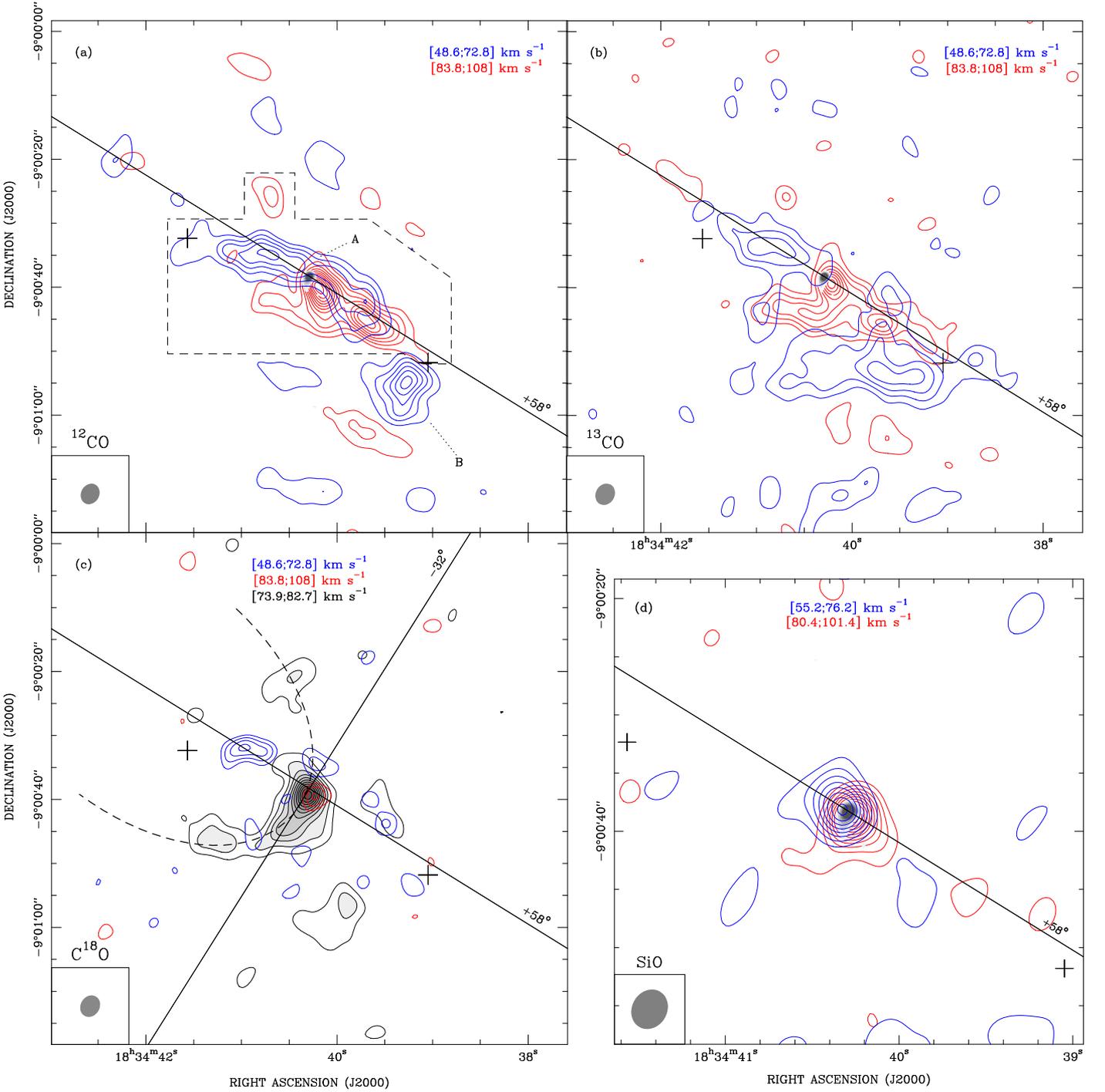}
\caption{SMA maps of the outflow emission from \G23.
\textbf{(a):} Integrated emission of the $^{12}$CO\,(2--1) line within the velocity ranges indicated in the top right.
The blue and red velocities are
symmetric w.r.t. the $V_{\rm sys}$ of the HMC. Contours start at
3\,$\sigma$ and increas in steps of 2\,$\sigma$ (corresponding to 3.8 and 2.6 Jy~beam$^{-1}$~km~s$^{-1}$,
for the red- and blue-shifted wings, respectively).
At the center of the field is the 1.3~mm continuum map (greyscale). The NE--SW line draws the direction of the outflow
inferred from the analysis of the HMC tracers (at $+58\degr$). Labels A and B indicate, respectively, the HMC and the
another CO clump probably lying along the line of sight (see Sect.~\ref{outdis}). The two plus signs mark the reference
points for the computation of the outflow length (details in Sect.~\ref{out}). The dashed pattern outlines the region
over which emission was averaged to obtain the spectra in Fig.~\ref{outspec}. The SMA synthesized beam is shown in the lower left corner.
\textbf{(b):} Same as panel (a), for the $^{13}$CO\,(2--1) line emission. Contours start at 3\,$\sigma$ and increase in steps of
2\,$\sigma$ (corresponding to 0.61 and 0.82 Jy beam$^{-1}$ km~s$^{-1}$, for the red- and blue-shifted emission, respectively).
\textbf{(c):} Same as panel (a), for the C$^{18}$O\,(2--1) line emission. The grey contours are a map of the C$^{18}$O bulk
emission integrated over the FWHM of the CH$_3$CN lines. The NW--SE line indicates the direction of the major axis of the HMC and the dashed
pattern mark the putative border of the NE outflow lobe, partially traced by the C$^{18}$O bulk emission. Contours start at 3\,$\sigma$ and increase in steps of 2\,$\sigma$ (0.59
Jy~beam$^{-1}$~km~s$^{-1}$) for the bulk emission, and of 1\,$\sigma$ for the red-shifted (0.28~Jy~beam$^{-1}$~km~s$^{-1}$) and blue-shifted
(0.33~Jy~beam$^{-1}$~km~s$^{-1}$) emission.
\textbf{(d):}  Similar to panel-(a) but for the SiO\,(5--4) line emission and with a zoom of two times the field of view.
Contours start at 3\,$\sigma$ and increase in steps of 2\,$\sigma$ (0.64 and 0.54~Jy~beam$^{-1}$~km~s$^{-1}$ for the red- and
blue-shifted emission, respectively).}
\label{outpuz}
\end{figure*}

\begin{figure}
\centering
\includegraphics [angle=-90.0,scale=0.5]{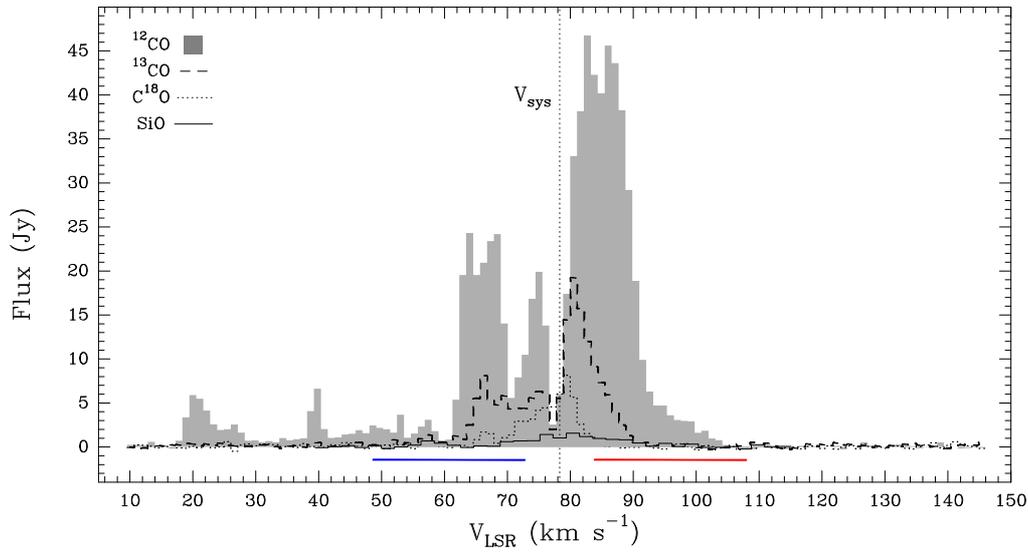}
\caption{Spectra of the outflow tracers, $^{12}$CO\,(2--1), $^{13}$CO\,(2--1), C$^{18}$O\,(2--1), and SiO\,(5--4) obtained
by averaging the emission inside
the dashed box in Fig.~\ref{outpuz}a. The ranges of integration used for the CO lines are marked along the x-axis
(blue and red lines); the rest velocity of the CH$_3$CN\,(12--11) lines is also shown ($\rm V_{sys}$).}
\label{outspec}
\end{figure}

\begin{figure*}
\centering
\includegraphics [angle=0.0,scale=0.9]{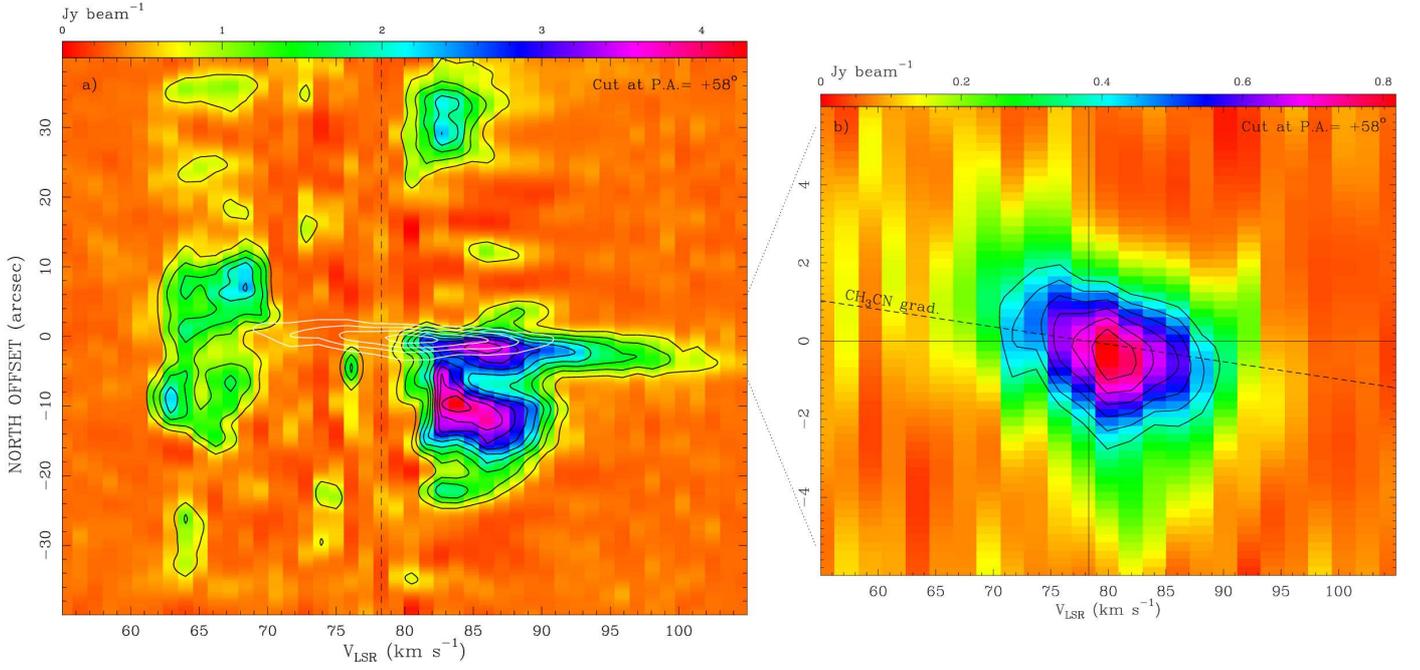}
\caption{\textit{p-v} plot of the outflow tracers $^{12}$CO\,(2--1) and SiO\,(5--4), along
a cut with P.A.=+58\degr\ (cf. Fig.~\ref{outpuz}a and \ref{outpuz}d). The vertical line marks the $V_{\rm sys}$.
\textbf{(a)} The $^{12}$CO emission is drawn with colors (color scale on top) and black contours, from 10\%
of the peak emission increasing by steps of 10\%. A map of the SiO emission is overlaid in white contours
starting from 30\% of the peak emission and increasing by 20\% steps.
\textbf{(b)} Close up view of the SiO {\it p-v} plot within 6\arcsec\ from the HMC. The SiO emission is drawn with
colors (color scale on top) and black contours, from 40\% of the peak emission increasing by steps of 10\%.
The dashed line denotes the velocity trend obtained from the CH$_3$CN emission (Fig.~\ref{velpuz}c).}
\label{outpv}
\end{figure*}

\begin{figure*}
\centering
\includegraphics [angle=0.0,scale=1.3]{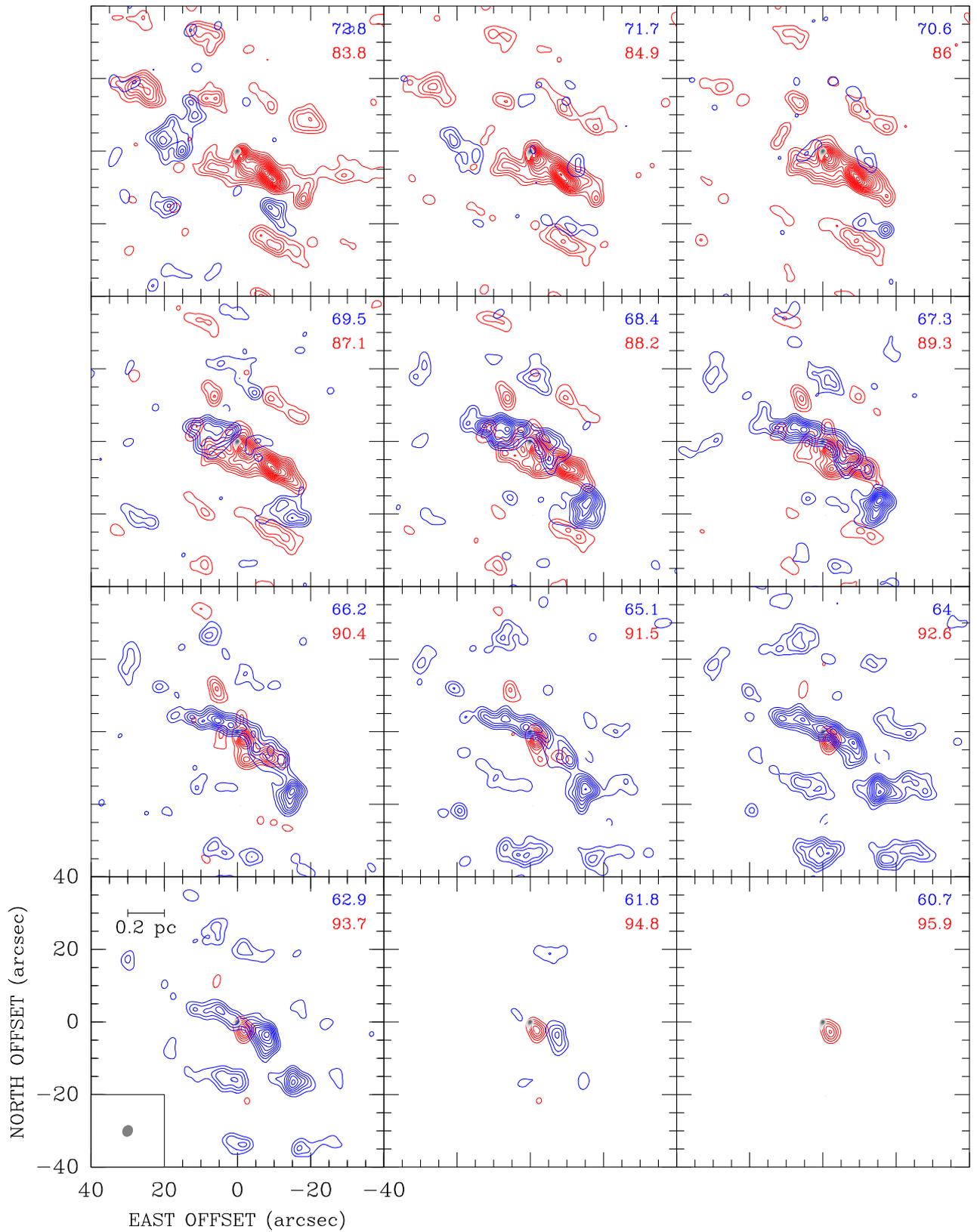}
\caption{Channel maps of the  $^{12}$CO\,(2--1) line emission observed with the compact configuration of the SMA. Each
box contains pairs of maps corresponding to the blue- and red-shifted emission at the same velocity offset
from $V_{\rm sys}$. The corresponding LSR velocities (km~s$^{-1}$) are indicated in the top right of each box. Contours start
from a 3\,$\sigma$ level of 0.4~\Jyb\ for both wings and increase in steps of 2\,$\sigma$. In the bottom left panel,
a linear scale and the synthesized beam are shown. The map of the 1.3~mm dust continuum (grey contours) is shown in overlay.}
\label{12cube}
\end{figure*}

\begin{figure*}
\centering
\includegraphics [angle=0.0,scale=0.9]{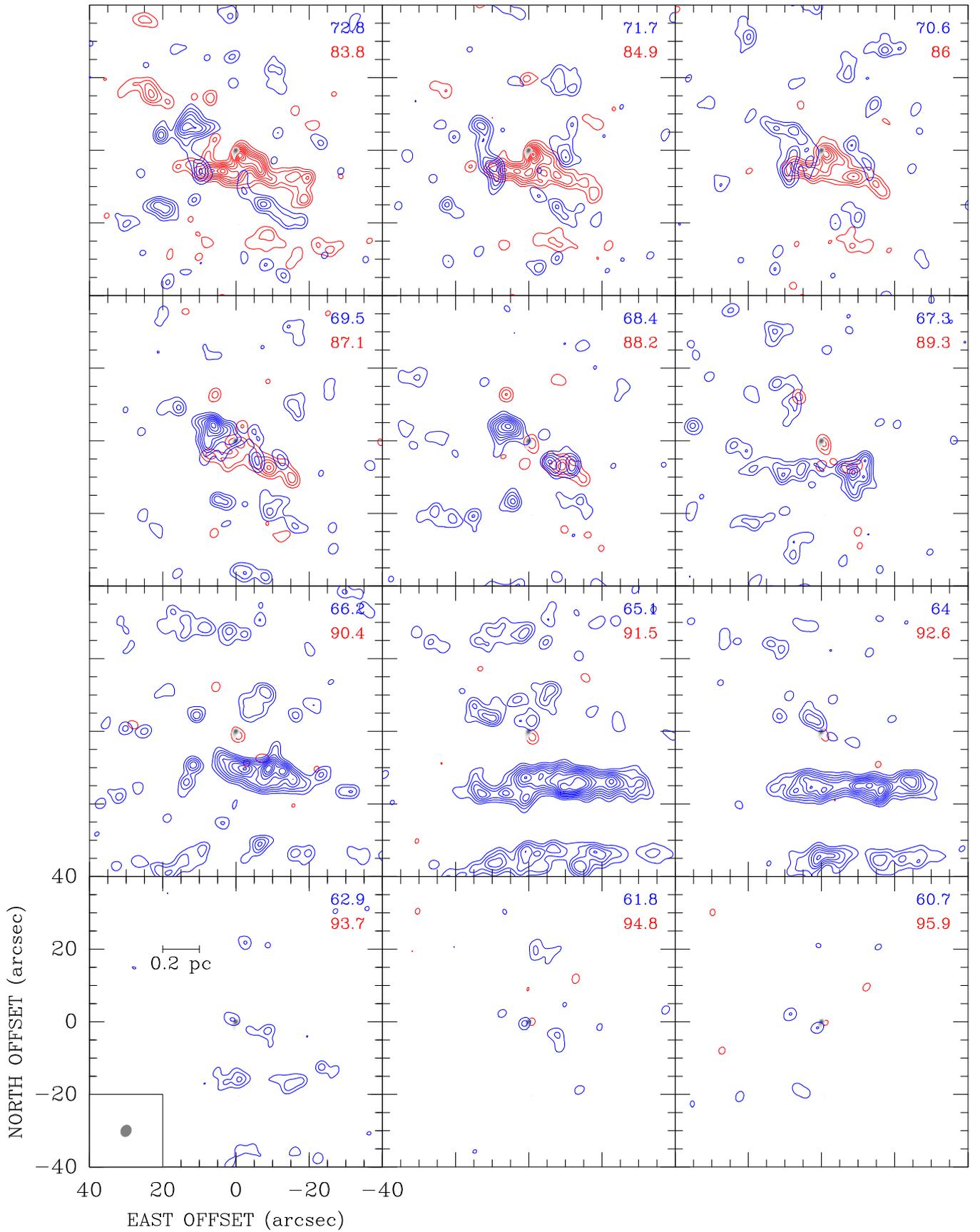}
\caption{Same as Fig.~\ref{12cube} but for the $^{13}$CO\,(2--1) line emission. Contours start from
a 3\,$\sigma$ level of 0.14~\Jyb\ and increase in steps of 2\,$\sigma$.}
\label{13cube}
\end{figure*}

\begin{figure*}
\centering
\includegraphics [angle=0.0,scale=0.95]{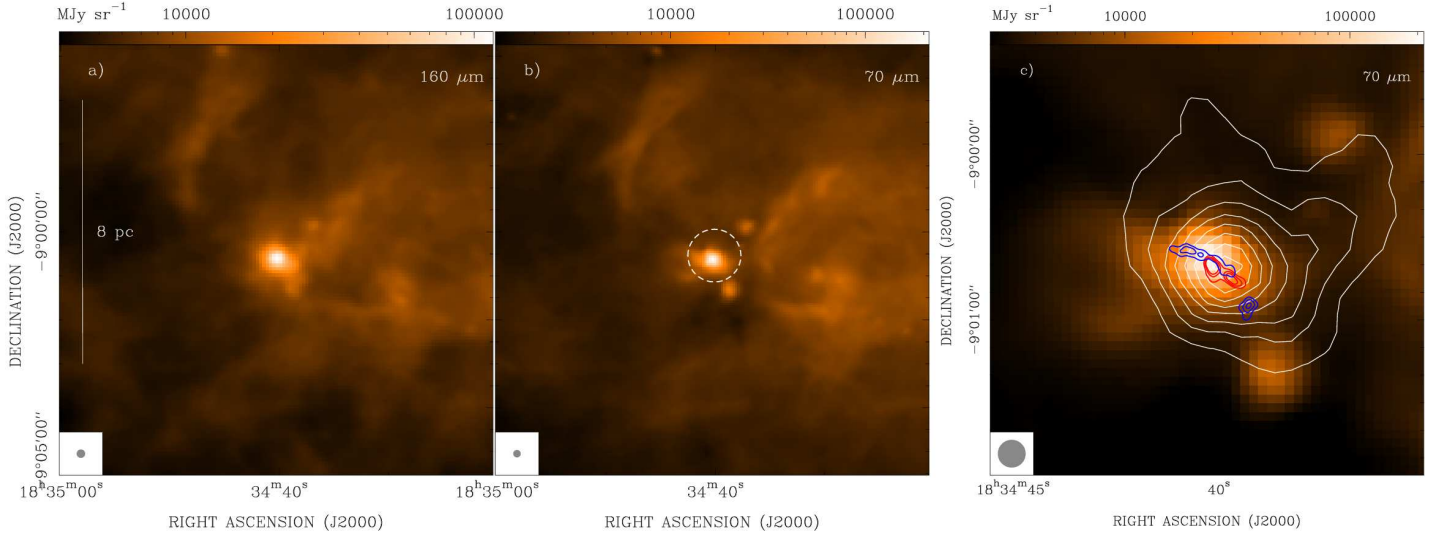}
\caption{Parsec-scale structure of the \G23\ star forming region obtained at $160~\mu$m
and $70~\mu$m (color log.-scale on top of each panel), and at $870~\mu$m (grey contours, right panel).
All plots are centered on the HMC, as done in Figs.~\ref{corepuz} and~\ref{outpuz}.
\textbf{(a)} Herschel view of the region at $160~\mu$m within $\pm$4.9\arcmin\ from the HMC. The linear scale
is drawn in the left side of the plot.
\textbf{(b)} Similar to the left panel but for the $70~\mu$m continuum emission. The white dashed circle
marks the region inside which the fluxes in Table~\ref{tsed} have been measured.
\textbf{(c)} Similar to panel~(b) but showing a close-up view of the region around the HMC.
The white contours are the ATLASGAL map at $870~\mu$m starting from 20\% of the peak emission and increasing
by steps of 10\%. Superimposed are also the contours of the $^{12}$CO\,(2--1) outflow emission of
Fig.~\ref{outpuz}a starting from a 7\,$\sigma$ level (blue and red contours).
For each plot, the beam size of the Herschel images is shown in the bottom left corner.}
\label{irplots}
\end{figure*}

\begin{figure*}
\centering
\includegraphics [angle=0.0,scale=0.7]{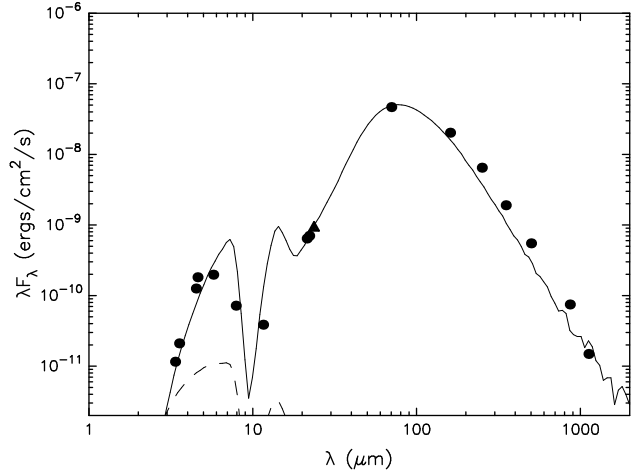}
\caption{Spectral energy distribution of \G23\ from 3.4~$\mu$m to 1.1~mm.
Dots (triangle) indicate measurements (lower limit) from different telescopes (see Table~\ref{tsed}).
The SED was fitted with the radiative transfer model developed by \citet[][and available as a fitting
tool at  http://caravan.astro.wisc.edu/protostars/]{Robitaille2007}. The solid black line indicates the
best-fitting model.}
\label{sed}
\end{figure*}

The velocity field in the HMC can be further investigated by means of the CH$_3$CN lines. As outlined in Section~\ref{core}, in
Fig.~\ref{velpuz} one can identify two velocity gradients, one in the direction of the equatorial plane of the flattened
HMC with increasing velocities to the SE (P.A.$\simeq$\,$-32\degr$), and the other along the perpendicular direction
(i.e., in the outflow direction at P.A.=$58\degr$) with velocities increasing to the SW (Figs.~\ref{velpuz}c and~\ref{velpuz}d).
Since there is no evidence for an outflow along the NW--SE direction (Sect.~\ref{outdis}), a possible interpretation
of this velocity field is that the CH$_3$CN gas is tracing {\it both} expansion along the outflow direction {\it and} rotation
perpendicular to it (see also Fig.~\ref{corepuz}b). As explained in Sect.~\ref{ch3cn}, the CH$_3$CN velocity gradient along the outflow
direction (see Fig.~\ref{velpuz}c) is well reproduced by a linear fit (corr. coeff.~0.8) and agrees with the velocity
gradient detected in the SiO\,(5--4) emission (Fig.~\ref{outpv}b) and that outlined by the methanol masers in
the \textit{p-v} plot of Fig.~\ref{velpuz}c (red triangles; \citealt{Sanna2010}).
These results suggest that, even inside a radius of $\sim$0\farcs25 from the center of the HMC, the CH$_3$CN emission is
affected by the outflow expansion. A similar velocity gradient is seen also
along the direction perpendicular to the outflow (Fig.~\ref{velpuz}d). The presence of two,
orthogonal, velocity gradients with similar slopes explains why the CH$_3$CN\,(6--5) maps by \citet[their Fig.~11]{Furuya2008}
did not reveal a clear velocity gradient, as their observations had a three times less resolution than ours.
Beyond the FWHM of the CH$_3$CN lines (red and blue dashed lines in Fig.~\ref{velpuz}d), the emission along a P.A.=\,$-32\degr$
appears to peak at zero offset, namely close to the outflow axis. This behavior may indicate that the high
velocity emission is dominated by gas entrained in the outflow, but it is also expected if the gas is undergoing Keplerian rotation. To
describe this scenario, we plot in Fig.~\ref{velpuz}d the velocity pattern corresponding to Keplerian rotation about a $\rm 19~M_\odot$
star, where the inclination of the disk with respect to the line-of-sight ($30\degr$; see Sect.~\ref{hmcdis}) has been taken into
account. While the Keplerian profile appears consistent with our data points, it is not sufficient to claim
the existence of a Keplerian disk. However, the presence of a velocity gradient inside the line FWHM seems clear
(Fig.~\ref{velpuz}b), and it is very reasonable that such a gradient could be due to rotation
(albeit not necessarily Keplerian) about an equilibrium mass on the order of $\rm 19~M_\odot$, in agreement with that estimated
with the CH$_3$OH maser spots (see above). In Fig.~\ref{velpuz}d, we show indeed that the maser points lying within a distance
of 0.25~arcsec from the dust continuum peak (red triangles) agree well with the peak distribution of the CH$_3$CN emission.

With this in mind, in the following we will adopt the working hypothesis that the HMC contains a flattened structure
whose rotation, about an axis close to $58\degr$ on the plane of the sky, implies an equilibrium mass of $\sim$19~$M_\odot$.

One may wonder what fraction of this mass is contributed by the gas lying inside $0\farcs25$ (or 1150~AU). The gas mass
can be estimated from the dust continuum emission (e.g., \citealt{Hildebrand1983}), assuming thermal equilibrium between gas and dust at a
temperature of 195~K, derived from the CH$_3$CN lines (see Table~\ref{tch3cn}). Our calculation uses a dust absorption
coefficient of 0.8~cm$^2$~g$^{-1}$ at the observing frequency and a standard gas-to-dust mass ratio of~100
\citep{Ossenkopf1994}. Inside a radius of 0\farcs65 or 3000~AU, corresponding to the region where both the CH$_3$CN lines
and the dust emission are detected with the combined array, we estimate a mass of $\rm 8.5\pm0.8~M_{\odot}$ and a mean
column density of $\rm (4.9 \pm 0.4) \times 10^{23}~cm^{-2}$. The quoted errors correspond to the $5\,\sigma$ uncertainty
of the flux measurement, although we stress that the main source of uncertainty comes from
the poorly known dust properties. Even assuming a density profile as steep as $n\propto R^{-2}$, the gas mass inside
1150~AU cannot be greater than $\rm 3.3~M_\odot$. We conclude that the stellar component must dominate the
$\rm \sim19~M_\odot$ estimated above.

\subsubsection{Stability of the rotating HMC}

It is also worth computing the mass in the envelope beyond the 3000~AU radius. For this purpose, we consider the
difference between the continuum flux measured with the compact SMA configuration and that measured with the combined
configuration, and assume a gas and dust temperature equal to that obtained by \citet{Codella1997} from the NH$_3$
lines (58~K). This choice is justified by the fact that ammonia observations were sensitive to more extended structures
than those imaged in our CH$_3$CN maps, similar to the largest scale imaged with the SMA for the dust emission (see
Sect.~\ref{hmcdis}). We thus estimate a mass of $12 \pm 4~M_{\odot}$ for the outer regions of the envelope, comparable
to that enclosed inside 3000~AU and to the putative stellar mass. An envelope whose mass is on the same order of that
of the central star(s) is expected to develop instabilities (see, e.g., \citealt{Cesaroni2007} and references therein).
It is hence of interest to verify whether or not the flattened, rotating structure observed in \G23\ is stable with
respect to axisymmetric gravitational disturbances, assuming that we can approximate the system as a thin
disk. Following the simplified approach outlined in \citet{Cesaroni2007}, we estimate the stability parameter $Q$
\citep{Toomre1964} at the maximum radius of emission for both the inner (inside 3000~AU) and outer (between 3000 and
8000~AU) regions of the HMC. The $Q$ parameter is proportional to the product of the sound speed and the angular velocity at a
given radius and inversely proportional to the disk surface density at the same radius. The condition of instability
is $Q<1$, which implies that a disk tends to form massive condensations.
At the inner radius of $\rm 3000~AU$, we estimate $Q\simeq9$, by replacing the sound speed with the linewidth of the
CH$_3$CN lines (${\rm FWHM}/\sqrt{8\ln2}$). At the outer radius of 8000~AU, using the NH$_3$ linewidth (from Table~8 of
\citealt{Codella1997}) and the mean surface density computed from the mass of $\rm 12~M_{\odot}$, we obtain a smaller
value, $Q\simeq3.5$. This estimate has to be taken as an upper limit because the NH$_3$ linewidth was measured at the
peak position of the VLA map. Note that at both radii the main support to gravitational instabilities is provided by turbulence,
which dominates the linewidth, since thermal broadening accounts for less than 1~\kms. While the error on these values of $Q$
is certainly large, the result that $Q$ decreases with distance from the HMC center appears more robust, because most sources
of uncertainties equally affect both estimates of $Q$. We thus conclude that the observed rotating structure appears to be
progressively more unstable at larger radii.

\subsection{Outflow morphology and dynamics}\label{outdis}

In Fig.~\ref{outpuz}a, we show that the $^{12}$CO emission extends along the NE--SW direction up to a distance of $\sim0.5$~pc from
the center of the HMC, with the red-shifted emission located to the SW. The $^{12}$CO outflow mass computed from the
$J$=\,2--1 transition is a factor $\sim 8$ less than that estimated by \citet{Furuya2008} from the $J$=\,1--0 transition
(after correcting for the new distance adopted by us). We believe that this discrepancy is compatible with the large
uncertainties of the $^{12}$CO\,(1--0) measurements (see Table~7 of \citealt{Furuya2008}), as well as to contamination by
extended emission in the $^{12}$CO\,(1--0) maps. For this reason, we will rely upon our estimates in the following discussion.

On a region four times smaller than that traced by $^{12}$CO and $^{13}$CO, the SiO\,(5--4) line emission traces a bipolar
outflow in the same direction as the $^{12}$CO\,(2--1) emission, with the blue- and red-shifted lobes located to the NE
and SW, respectively (Fig.~\ref{outpuz}d). We stress that the direction of the outflow (P.A.=$58\degr$) coincides with the
rotation axis of the flattened HMC (Fig.~\ref{corepuz}). It is also worth noting that the outflow direction agrees with both
the orientation and proper motion of the H$_2$O maser jet detected by \citet[their Fig.~5b]{Sanna2010}. The maser emission
arises within 2000~AU from the dust continuum peak detected with the SMA. This evidence supports the existence of a powering
source at the center of the dusty core which dominates the outflow dynamics at all scales.

The CO outflow lobes appear very collimated with a width-to-length ratio $\sim0.3$, which suggests that the outflow axis
lies close to the plane of the sky, in agreement with the small inclination ($30\degr$) of the disk-like structure of the HMC
with respect to the line-of-sight (see Sect.~\ref{hmcdis}). This is also consistent with the detection of some faint blue-shifted
emission from the red lobe, and vice versa, as expected when the lobes are partly crossing the plane of the sky.  In
Table~\ref{tout}, we also report the parameters of the outflow corrected for an inclination angle of $30\degr$.

Unlike the C$^{18}$O\,(1--0) line, the C$^{18}$O\,(2--1) line emission peaks at the position of the HMC with a
fainter tail extending to the SE (Fig.~\ref{outpuz}c), towards the peak of the $J$=\,1--0 line (see Fig.~6c of \citealt{Furuya2008}).
The overall bulk emission of the C$^{18}$O gas to the NE presents a curved morphology, outlined by the dashed pattern in
Fig.~\ref{outpuz}c. While the faint, high-velocity C$^{18}$O emission is aligned with the outflow axis,
the quiescent C$^{18}$O gas (within the FWHM of the CH$_3$CN emission) might arise from the outer edges of the
blue-shifted outflow lobe.

The channel maps reveal a clumpy structure along the outflow axis (Fig.~\ref{12cube}), possibly related to different
ejection episodes. Such a clumpyness is also visible in the \textit{p-v} plot in Fig.~\ref{outpv}, where one may also
appreciate the lack of emission close to the systemic velocity, likely due to extended structures filtered out by the
interferometer. In contrast, the SiO emission in Fig.~\ref{outpv} is detected at all velocities, although the peak appears offset with
respect to the systemic velocity. This agrees with the idea that the SiO emission originates mostly
from the high-velocity gas in the post-shock region, where the SiO molecules are believed to form (see Sect.~\ref{sio}).

A remarkable fact is that the blue-shifted emission in all CO isotopomers is significantly fainter
and much more patchy than the red-shifted emission. This is clearly seen in the channel maps as well as in the
mean spectra in Fig.~\ref{outspec}, which reveal the presence of multiple velocity components below 80~\kms,
likely due to unrelated molecular clouds along the line of sight to \G23. In contrast, the red-shifted
emission is much more coherent and the corresponding spectral profile smoother. As expected, the contamination by
spurious components at blue-shifted velocities becomes much less pronounced in the optically thinner C$^{18}$O line and is
basically absent in the SiO transition. We conclude that most of the features observed in the blue-shifted channels
are due to molecular gas along the line of sight, which has no relationship with the region of interest for us. This explains
the presence of prominent blue-shifted emission on the same side as the red lobe, especially the elongated $^{13}$CO structure
centered at ($-10\arcsec$,$-15\arcsec$) and spanning a velocity range from 64 to 66.2~\kms\ (see Fig.~\ref{13cube}). Note that
the same structure, named clump~B in Sect.~\ref{coiso}, looks fainter and more fragmented in the corresponding
$^{12}$CO maps (Fig.~\ref{12cube}), which confirms that the emission originates from an extended cloud along the line
of sight, partly resolved out in the optically thicker transition.

In the calculation of the outflow parameters we have payed attention to select only the emission most likely associated
with the outflow, both in space and velocity. We are thus confident that the parameters listed in Table~\ref{tout} are
reliable in this respect. Despite the large uncertainty on the SiO abundance, which in turn affects the reliability
of the corresponding outflow parameters, we note that the latter differ only by a factor $\sim$4 from those obtained
from the CO line. All these facts make us confident that our estimates are reliable.

Finally, we note that a momentum rate of $6\times10^{-3}$~M$_\odot$~km~s$^{-1}$~yr$^{-1}$, determined from the
$^{12}$CO\,(2--1) emission after correcting for the inclination angle of the outflow, is in good agreement with
the bolometric luminosity, according to a well known empirical relation (see Fig.~4 of \citealt{Beuther2002}). This result
lends further support to our estimate of the luminosity (and mass) of the embedded star.

\subsection{Nature of the embedded source(s) in the HMC}

In Sect.~\ref{hmcdis} we have proposed a scenario where the HMC is rotating down to scales on the order of 1000~AU,
and the corresponding equilibrium mass is $\rm \la19~M_\odot$. We have also shown that most of this mass must be in the
form of (proto)stars. In the following, we want to set constraints on the nature of these objects using the information
obtained from the IR and free-free radio emission.

In Sect.~\ref{SED}, we concluded that in all likelihood the HMC is dominating the IR flux of the region (see
Fig.~\ref{irplots}) and hence is responsible for the bolometric luminosity of $\rm 3.9\times10^4~L_{\odot}$.
This conclusion is strengthen by a simple calculation of the HMC luminosity if it was in thermal equilibrium
at the temperature of the CH$_3$CN gas. By using $\rm T_{rot}=$\,195~K and assuming LTE inside the HMC, within
the radius of  CH$_3$CN emission ($\rm \la3000~AU$) the Stefan-Boltzmann law \textbf{($\rm L \propto T^4$)}
yields a bolometric luminosity of $\sim 5\times10^4~L_{\odot}$, which is in remarkable agreement with the luminosity
obtained from the measured SED. For a ZAMS star, the measured luminosity is comparable to the expected luminosity
($\rm \sim5\times10^4~L_\odot$) of a $\rm 19~M_\odot$ star (e.g., Eq.~13 of \citealt{Hosokawa2010}), suggesting that
a single object with a O9.5 spectral type could justify the observed luminosity. However, as noted by \citet{Sanna2010},
such a star should produce much more Lyman continuum photons than those estimated from the measured radio continuum
($\sim3\times10^{45}$~s$^{-1}$). In fact, the free-free emission, which peaks at the center of the HMC
(\citealt{Sanna2010}, their Table~1 and Fig.~4), corresponds to photo ionization by a ZAMS star with a B1 spectral
type, namely a ZAMS luminosity an order of magnitude less than the observed luminosity (e.g., \citealt{Schaller1992}).
How may one reconcile this evidence with the measured bolometric luminosity?

A first possibility is that ongoing accretion onto the star is quenching the formation of an \HII\ region (e.g.,
\citealt{Keto2002}), allowing only a small fraction of the Lyman continuum photons to escape through the outflow.
The measured free-free continuum would hence lead to an underestimate of the true Lyman photons and, hence, of
the stellar luminosity. Alternatively, the free-free emission could be tracing a thermal jet where the ionization is
due to shocks in the ejected material, as we have speculated in \citet{Sanna2010} on the basis of the
radio continuum morphology and spectral index between 3.6~and 1.3~cm. As a matter of fact, jets like these -- albeit
rare -- have been found associated with high-mass stars (e.g., \citealt{Anglada1996}).

Another explanation may be that we are dealing with a massive, bloated pre-ZAMS star as those modeled by
\citet{Hosokawa2010} in the presence of high accretion rates ($\rm \ga 10^{-3}~M_{\odot}~yr^{-1}$). Objects of
this type are much colder than a ZAMS star with the same luminosity and are hence very weak Lyman continuum emitters.
However, the accretion rate onto the star should be a few times less than the mass loss rate determined from the outflow
tracers (e.g., \citealt{Behrend2001}), which in our case implies a value $\rm \la10^{-4}~M_{\odot}~yr^{-1}$ (see Table~\ref{tout}).
Also, objects as massive as $\rm 19~M_\odot$ are not expected to have a pre-ZAMS phase (see Fig.~14 of \citealt{Hosokawa2010}).
Therefore, we consider this scenario quite unlikely.

Finally, one should consider the possibility that the total mass ($\rm 19~M_\odot$) is contributed by multiple lower-mass
stars, rather than a single O9.5 star. This hypothesis cannot be proved or ruled out by direct observations, as the angular
resolution of the available images is insufficient to resolve a tight binary or a multiple system in the HMC.
Splitting a mass of $\rm 19~M_\odot$ into smaller stars would reduce the Lyman continuum output significantly and make it
consistent with that estimated from the radio continuum. However, the bolometric luminosity would also decrease well below
the observed value  ($\rm L_{\star, ZAMS} \propto M_\star^2$), which appears inconsistent with the fact that most of the
observed emission in the infrared originates from the HMC (see Sect.~\ref{SED}).

In conclusion, we believe that, with the current data, the most reliable scenario for \G23\ is that of a deeply embedded
massive ($\la$19~$M_\odot$) ZAMS star, still undergoing accretion from the parental core, possibly surrounded by an accretion
disk, and powering a radio jet and a pc-scale outflow.

\section{Summary and conclusions}

We used the SMA to image various hot-core and outflow tracers from the
high-mass star forming region \G23\ at 1.3~mm. We have also combined the
Herschel/Hi-GAL data with ancillary archival data to reconstruct the SED
associated with the HMC in \G23\ (between~3.4~$\mu$m and~1.1~mm) and derive
its luminosity. In this paper, we have focused on a comparison of the gas
dynamics in the inner HMC, on scales $\rm \la 0.01~pc$, with respect to that
inferred from typical outflow tracers, on scales $\rm \ga 0.1~pc$.
Our main results can be summarized as follows:

\begin{enumerate}
\item We find that the HMC core is significantly flattened (width-to-length
ratio $\sim2$), which corresponds to an inclination of $30\degr$ under the
assumption that we are observing an oblate core tilted with respect to the
line of sight. Also, the size of the core appears to decrease with excitation
energy of the observed transition, suggesting the presence of a temperature
gradient due to embedded YSOs. From the mean CH$_3$CN spectrum within a
radius of $\sim$3000~AU from the HMC center, we infer a rotational temperature
of $\sim$195~K.
\item The CH$_3$CN emission appears to trace two distinct velocity fields,
one consistent with expansion along the outflow axis, the other suggesting
rotation about the same axis (possibly Keplerian), in good agreement with the (3-D)
velocity field traced by methanol masers. The central mass implied by the
rotational component is $\sim$19~$M_\odot$, after taking into account an
inclination of the rotation axis of $30\degr$. This value is consistent with the bolometric
luminosity of the \G23\ region ($\sim$$4\times10^4~L_\odot$) estimated by
fitting the SED.
\item We have estimated the Tommre $Q$ stability parameter at two different
radii in the putative disk, and find that $Q$ decreases
significantly from small to large radii, consistent with the disk being more
stable close to the central star(s), as expected.
\item The outflow emission imaged in the CO isotopomers and SiO(5--4) line
well matches the direction of the rotation axis (P.A.=\,$58\degr$) of the flattened HMC.
The outflow is well collimated with a width-to-length ratio for each lobe of about 0.3.
The energetics of the molecular outflow, such as its momentum rate of
$6\times10^{-3}$~M$_\odot$~km~s$^{-1}$~yr$^{-1}$, is in good agreement with that
expected for the measured bolometric luminosity.

\end{enumerate}

We conclude that the HMC in \G23\ is likely hosting an O9.5 (proto)star lying
at its center, powering the bipolar outflow observed on the parsec scale, and shows clear
proof of rotation up to a radius of $\sim$1000~AU. Higher angular
resolution observations are needed in order to confirm that the HMC is indeed
hosting a single massive star, to trace the inner rotation curve, and to find out
whether accretion is still going on inside the HMC.

\begin{acknowledgements}

Comments from an anonymous referee which helped improving our paper are gratefully acknowledged.
Financial support by the European Research Council for the ERC Advanced Grant GLOSTAR (ERC-2009-AdG, Grant Agreement no. 247078)
is gratefully acknowledged. This research made use of the myXCLASS program (https://www.astro.uni-koeln.de/projects/schilke/XCLASS),
which accesses the CDMS (http://www.cdms.de) and JPL (http://spec.jpl.nasa.gov) molecular data bases.
This publication makes use of data products from the Wide-field Infrared Survey Explorer, which is a
joint project of the University of California, Los Angeles, and the Jet Propulsion Laboratory/California Institute of
Technology, funded by the National Aeronautics and Space Administration.

\end{acknowledgements}

\bibliographystyle{aa}
\bibliography{asanna2511}


\clearpage

\clearpage

\clearpage

\clearpage

\clearpage

\clearpage

\begin{landscape}
\begin{table}
\centering
\caption{Summary of SMA observations (code 2009B--S032) \label{tobs}}
\begin{tabular}{c c c c c c c c c c c}

\hline \hline

& & & &\multicolumn{2}{c}{Freq.\,Cove.\,(GHz)} &  &  &  &  & \\
\cline{5-6} \\
Obs.\,Date & Array\,Conf. & R.A.\,(J2000) & Dec.\,(J2000) &  LSB & USB & Spec.\,Res.\,(MHz) & Bandpass\,Cal. & Phase\, Cal. & Flux\,Cal. & HPBW \\
 (1) & (2) & (3) & (4) & (5) & (6) & (7) & (8) & (9) & (10) & (11)  \\
\hline
2010 Mar 05 &   VEX   (6) & $18^h34^m40\fs29$ & $-09\degr00'38\farcs3$ & 216.9$-$220.9 & 228.9$-$232.9 & 0.81\tablefootmark{a} & J1924$-$292    & J1743$-$038\,J1911$-$201 & MWC349 & $0\farcs4$ \\
2010 Apr 28 & Compact (8) & $18^h34^m40\fs29$ & $-09\degr00'38\farcs3$ & 216.9$-$220.9 & 228.9$-$232.9 & 0.81\tablefootmark{a} & 3C273\,3C454.3 & J1743$-$038\,J1911$-$201 & Titan & $3\farcs5$ \\
\hline
\end{tabular}

\tablefoot{Column\,(1): observing dates. Column\,(2): array configurations and number of effective antennas. Columns\,(3)-(4): phase center
of the observations. Columns\,(5)-(6): approximate ranges of rest frequencies covered in the lower and upper sidebands. Column\,(7): spectral
resolution corresponding to a velocity width of 1.05~\kms. Columns\,(8)-(10): bandpass, phase, and absolute flux calibrators. Columns\,(11): approximate size of the beam obtained with a natural weighting at each configuration.\\
\tablefoottext{a}{The spectral resolution was improved on the range 220.51$-$220.73~GHz by a factor of 2 to better sample the
CH$_3$CN\,(12$_K$--11$_K$) lines (see Sect.\ref{smaobs}).}
}
\end{table}
\end{landscape}

\clearpage

\begin{table*}
\caption{Observed lines\label{tlines}}

\begin{tabular}{c c r}

\hline \hline

\multicolumn{1}{c}{$\nu$}  &  \multicolumn{1}{c}{Line} & \multicolumn{1}{c}{E$_{\rm low}$}  \\ 
\multicolumn{1}{c}{(GHz)}  &                           & \multicolumn{1}{c}{(K)}            \\ 

\hline
217.105 & SiO\,(5--4)                &  20.8  \\ 
219.560 & C$^{18}$O\,(2--1)          &   5.3  \\ 
220.399 & $^{13}$CO\,(2--1)          &   5.3  \\ 
220.476 & CH$_3$CN\,($12_8-11_8$)    & 515.0  \\ 
220.486 & CH$_3^{13}$CN($12_6-11_6$) & 313.9  \\ 
220.532 & CH$_3^{13}$CN($12_5-11_5$) & 235.8  \\ 
220.539 & CH$_3$CN\,($12_7-11_7$)    & 408.0  \\ 
220.570 & CH$_3^{13}$CN($12_4-11_4$) & 171.9  \\ 
220.594 & CH$_3$CN\,($12_6-11_6$)    & 315.3  \\ 
220.600 & CH$_3^{13}$CN($12_3-11_3$) & 122.2  \\ 
220.621 & CH$_3^{13}$CN($12_2-11_2$) &  86.7  \\ 
220.634 & CH$_3^{13}$CN($12_1-11_1$) &  65.3  \\ 
220.638 & CH$_3^{13}$CN($12_0-11_0$) &  58.2  \\ 
220.641 & CH$_3$CN\,($12_5-11_5$)    & 236.8  \\ 
220.679 & CH$_3$CN\,($12_4-11_4$)    & 172.6  \\ 
220.709 & CH$_3$CN\,($12_3-11_3$)    & 122.6  \\ 
220.730 & CH$_3$CN\,($12_2-11_2$)    &  86.8  \\ 
220.743 & CH$_3$CN\,($12_1-11_1$)    &  65.4  \\ 
220.747 & CH$_3$CN\,($12_0-11_0$)    &  58.3  \\ 
229.589 & CH$_3$OH\,($15_4-16_3$)E   & 363.4  \\ 
230.538 & $^{12}$CO\,(2--1)          &   5.5  \\ 

\hline
\end{tabular}

\tablefoot{Values of frequencies and lower energies for each molecular transition are reported
from the Cologne Database for Molecular Spectroscopy, CDMS \citep{Muller2005}.}\\

\end{table*}

\begin{landscape}
\begin{table}
\centering
\caption{\G23: core tracers \label{tcore}}

\begin{tabular}{c c c c c c c c c c c c c}

\hline \hline
       & & & & & & & \multicolumn{2}{c}{Peak\,\,Position} & & & \multicolumn{2}{c}{Deconv.\,\,Size}\\
 \cline{8-9} \cline{12-13} \\
  & Array & Weight &  \multicolumn{2}{c}{HPBW\, \& \, P.A.} & Image rms & $\rm \Delta V_{\rm LSR}$ &  R.A.\,(J2000) &  Dec.\,(J2000) & F$_{\rm peak}$  & F$_{\rm int}$ & $\theta_{M} \times \theta_{m}$ & P.A. \\
Tracer  &  & & ($''$) & ($\degr$) & (Jy beam$^{-1}$) & (km s$^{-1}$) &(h m s) & ($\degr$ $'$ $''$)& (Jy beam$^{-1}$) & (Jy) & ($''$)& ($\degr$) \\
\hline
Dust 1.3~mm & Compact & NA & $3.8 \times 3.3$ & -15 & 0.006 & ... & 18:34:40.284 ($\pm 0\farcs1$) & -9:00:38.314 ($\pm 0\farcs1$) & $0.192\pm0.010$ & 0.294 & $3.4\times1.9 $ & -16 \\
 & Comb. & NA & $0.97 \times 0.94$ & 25 & 0.004 & ... & 18:34:40.292 ($\pm 0\farcs05$) & -9:00:38.354 ($\pm 0\farcs05$) & $0.102 \pm 0.008$ & 0.209 & $1.20 \times 0.60$ & -31.7 \\
 & VEX   & UN & $0.44 \times 0.31$ & 38 & 0.001 & ... & 18:34:40.290 ($\pm 0\farcs01$) & -9:00:38.183 ($\pm 0\farcs01$) & $0.035 \pm 0.001$ & 0.040 & \multicolumn{2}{c}{unresolved} \\
& & & & & & & & & & & &  \\
 \multicolumn{13}{c}{\textbf{CH$_3$CN\,($12_k-11_k$) data cube}}  \\
$\rm K-ladder$\tablefootmark{a} & Comb. & NA & $1.12 \times 1.07$ & -15 & 0.03 & ... & 18:34:40.286  ($\pm 0\farcs02$) & -9:00:38.374 ($\pm 0\farcs02$) & 1.57 & ...  & ... & ... \\
& & & & & & & & & & & &  \\
 \multicolumn{13}{c}{\textbf{CH$_3$CN\,(12--11) moment-0 maps}}  \\
$\rm K=2$ & Comb. & UN & $0.72 \times 0.57$ & 35 & 0.16\tablefootmark{b} & 75.6--81 & 18:34:40.286 ($\pm 0\farcs02$) & -9:00:38.332 ($\pm 0\farcs02$) & ...  & ...  & $1.39 \times 0.68$ & -19 \\
$\rm K=3$ & Comb. & UN & $0.72 \times 0.57$ & 35 & 0.18\tablefootmark{b} & 75.6--81 & 18:34:40.288 ($\pm 0\farcs03$) & -9:00:38.397 ($\pm 0\farcs03$) & ...  & ...  & $1.18 \times 1.12$ & -50 \\
$\rm K=7$ & Comb. & UN & $0.72 \times 0.57$ & 35 & 0.14\tablefootmark{b} & 75.6--81 & 18:34:40.287 ($\pm 0\farcs03$) & -9:00:38.293 ($\pm 0\farcs03$) & ...  & ...  & $0.65 \times 0.48$ & -49 \\
$\rm K=3~red$  & Comb. & UN & $0.72 \times 0.57$ & 35 & 0.06\tablefootmark{b} &   81--83   & 18:34:40.294 ($\pm 0\farcs04$) & -9:00:38.300 ($\pm 0\farcs04$) & ...  & ...  & ... & ... \\
$\rm K=3~blue$ & Comb. & UN & $0.72 \times 0.57$ & 35 & 0.04\tablefootmark{b} & 73.6--75.6 & 18:34:40.270 ($\pm 0\farcs04$) & -9:00:38.230 ($\pm 0\farcs04$) & ...  & ...  & ... & ... \\
& & & & & & & & & & & &  \\
 \multicolumn{13}{c}{\textbf{CH$_3$OH\,(15$_4$--16$_3$)~E maps}}  \\
CH$_3$OH\tablefootmark{c} & Comb. & R0 & $0.71 \times 0.58$  & 47 & 0.03 & ... & 18:34:40.291 ($\pm 0\farcs02$) & -9:00:38.250 ($\pm 0\farcs02$) & $0.264 \pm 0.013$ & 0.520 & $0.85 \times 0.34$ & -33 \\
CH$_3$OH~red  & Comb. & R0 & $0.71 \times 0.58$  & 47 & 0.07\tablefootmark{b} & 79.4--81.6 & 18:34:40.289 ($\pm 0\farcs04$) & -9:00:38.440 ($\pm 0\farcs04$) & ...  & ...  & ... & ... \\
CH$_3$OH~blue & Comb. & R0 & $0.71 \times 0.58$  & 47 & 0.08\tablefootmark{b} & 73.9--76.1 & 18:34:40.289 ($\pm 0\farcs04$) & -9:00:38.160 ($\pm 0\farcs04$) & ...  & ...  & ... & ... \\
\hline
\end{tabular}

\tablefoot{Column\,(1): molecular tracer. Column\,(2): array configurations; ``Comb.'' stands for the combined imaging of the
Compact and VEX data. Columns\,(3)-(5): image weighting, synthesized beam, and position angle (east of north) of each map.
Column\,(6): 1\,$\sigma$ rms of the map. Column\,(7): integrated velocity ranges (see Sect.~\ref{core}).
Columns\,(8)-(9): for maps of dust emission and lines measured within the bulk emission (Sect.~\ref{core}), the peak position was derived
by an elliptical Gaussian fitting (with fitted errors). For those line maps integrated over the red- and blue-shifted emission, we give
the position of the peak. Column\,(10): peak brightness derived by gaussian fitting (with fitted errors).
Column\,(11): total integrated flux of the emission within the 5\,$\sigma$ contour level.
Columns\,(12)-(13): deconvolved major, minor axes and position angle of the emission derived by gaussian fitting.\\
\tablefoottext{a}{The absolute peak position and its intensity correspond to the component $K$=1.}
\tablefoottext{b}{Units of Jy~beam$^{-1}$~km~s$^{-1}$.}
\tablefoottext{c}{Channel map at the systemic velocity of the HMC (78.3~km~s$^{-1}$).}
}
\end{table}
\end{landscape}

\begin{table}
\caption{Results of the $\rm CH_3CN\,(12_K-11_K)$ spectra analysis \label{tch3cn}}
\begin{tabular}{c c c c}
\hline \hline
K-component & V$_{\rm LSR}$ &  $\rm \Delta v$ &  $\int T_{B} dv$ \\
            & (km s$^{-1}$) & (km s$^{-1}$)   &  (K km s$^{-1}$) \\
\hline
K = 0 & $78.34 \pm 0.04$ & $ 9.43 \pm 0.05$ & $ 62 \pm 1 $   \\
K = 1 & & &  $ 63 \pm 1 $  \\
K = 2 & & &  $ 66.4 \pm 0.8 $  \\
K = 3 & & &  $ 69.2 \pm 0.8 $ \\
K = 4 & & &  $ 45.7 \pm 0.8 $ \\
\multicolumn{4}{c}{XLCASS synthetic spectrum:}  \\
\multicolumn{4}{l}{Source size (diameter): 1.2~arcsec} \\
\multicolumn{4}{l}{Rotational Temp.: 195~K} \\
\multicolumn{4}{l}{Column density: $5.1\times10^{16}$~cm$^{-2}$} \\
\hline
\end{tabular}
\tablefoot{~Properties of the CH$_3$CN\,$(12_K-11_K)$ line emission. We list the centroid velocity,
line width (FWHM), and line intensities from the multiple-Gaussian fitting, and the best-fit parameters
obtained with XCLASS (details on Sect.~\ref{core} and Fig.~\ref{ch3cnspec}).}

\end{table}

\clearpage

\begin{table*}
\caption{\G23: properties of the molecular outflow}
\label{tout}
\begin{tabular}{c c c c c c c c c c}

\hline \hline


       & & R & t$_{\rm dyn}$ & M$_{\rm out}$ & $\rm \dot{M}_{\rm out}$ & p & $\rm \dot{p}$ & E$_{\rm mec}$ & L$_{\rm mec}$  \\ 
Tracer & Lobe&(pc) & (yr) & (M$_{\odot}$) & (M$_{\odot}$ yr$^{-1}$) & (M$_{\odot}$ km~s$^{-1}$) & (M$_{\odot}$ km~s$^{-1}$~yr$^{-1}$) & ($10^{46}$~erg) & (L$_{\odot}$) \\  

\hline
$^{12}$CO & Red--A  & 0.51 & $ 3.4 \times 10^4 $ & 2.5 & $ 0.8 \times 10^{-4} $ & 37.5 & $ 1.1 \times 10^{-3} $ & 0.6 & 1.4 \\ 
          & Blue--A & 0.44 & $ 2.9 \times 10^4 $ & 1.4 & $ 0.5 \times 10^{-4} $ & 20.5 & $ 0.7 \times 10^{-3} $ & 0.3 & 0.9 \\ 
SiO       & Red--A  & 0.13 & $ 5.5 \times 10^3 $ & 1.0 & $ 1.8 \times 10^{-4} $ & 23.1 & $ 4.2 \times 10^{-3} $ & 0.5 & 7.9 \\ 
          & Blue--A & 0.12 & $ 5.1 \times 10^3 $ & 0.6 & $ 1.3 \times 10^{-4} $ & 14.9 & $ 2.9 \times 10^{-3} $ & 0.3 & 5.5 \\ 
 &  &  &  &  &  &  &  &  \\
       & & $\rm R \times \frac{1}{cos~30\degr} $ & $\rm t_{dyn} \times tan~30\degr$ & M$_{\rm out}$ & $\rm \dot{M}_{\rm out} \times cot~30\degr$ &
$\rm p \times \frac{1}{sin~30\degr} $ & $\rm \dot{p}  \times \frac{cos~30\degr}{sin^2~30\degr} $ & $\rm E_{mec} \times \frac{1}{sin^2~30\degr}$ &
L$_{\rm mec} \times \frac{cot~30\degr}{sin^2~30\degr}$ \\
 &  &  &  &  &  &  &  &  \\
$^{12}$CO &  & 0.55 & $ 1.8 \times 10^4 $ & 4.0 & $ 2.2 \times 10^{-4} $ & 116 & $ 6.2 \times 10^{-3} $ & 3.6 & 16 \\ 
SiO       &  & 0.14 & $ 3.1 \times 10^3 $ & 1.6 & $ 5.4 \times 10^{-4} $ &  76 & $ 2.4 \times 10^{-2} $ & 3.2 & 92 \\
 &  &  &  &  &  &  &  &  \\
\hline
\end{tabular}

\tablefoot{Properties of the molecular outflow estimated from the integrated emission in Fig.~\ref{outpuz} (see details in
Sect.~\ref{out}). Values not corrected for inclination are given for each lobe separately and are the following:
the maximum length of the outflow lobe (R); the dynamical time scale (t$_{\rm dyn}$); the outflow lobe mass and the
associated outflow rate (M$_{\rm out}$ \& $\rm \dot{M}_{\rm out}$); the momentum and mechanical force (p \& $\rm \dot{p}$);
the mechanical energy and luminosity (E$_{\rm mec}$ \& L$_{\rm mec}$).
The last two lines give the total outflow parameters, corrected for an inclination angle of $30\degr$
with respect to the plane of the sky.}
\end{table*}

\clearpage

\begin{table*}
\caption{\G23: IR spectral energy distribution within the dashed box of Fig.~\ref{irplots}b \label{tsed}}

\begin{tabular}{r r c}

\hline \hline

\multicolumn{1}{c}{$\lambda$}  &  \multicolumn{1}{c}{F$_{\rm int}$}  & Instrument \\
\multicolumn{1}{c}{($\mu$m)}   &  \multicolumn{1}{c}{(Jy)}           &            \\

\hline

3.4 & 0.013  &  WISE            \\ 
3.6 & 0.025  &  Spitzer/GLIMPSE \\ 
4.5 & 0.19   &  Spitzer/GLIMPSE \\ 
4.6 & 0.28   &  WISE            \\ 
5.8 & 0.38   &  Spitzer/GLIMPSE \\ 
8.0 & 0.19   &  Spitzer/GLIMPSE \\ 
12 & 0.15    &  WISE            \\ 
21 & 4.6     &  MSX             \\ 
22 & 5.2     &  WISE            \\ 
24 & $>7.2$  &  Spitzer/MIPSGAL \\ 
70 & 930     & Herschel/Hi-GAL  \\ 
160& 1090    & Herschel/Hi-GAL  \\ 
250& 540     & Herschel/Hi-GAL  \\ 
350& 220     & Herschel/Hi-GAL  \\ 
500& 92      & Herschel/Hi-GAL  \\ 
870& 22      & APEX/ATLASGAL    \\ 
1100 & 5.6   & CSO/BGPS         \\ 
\hline
\end{tabular}


\end{table*}

\end{document}